\documentclass[11pt,letterpaper]{article}
\usepackage[top=0.7in,left=0.7in,footskip=0.7in,marginparwidth=2in]{geometry}

\usepackage{lineno}
\usepackage[utf8]{inputenc}

\usepackage{cite}

\usepackage{nameref,hyperref}

\usepackage{setspace}
\usepackage{microtype}
\DisableLigatures[f]{encoding = *, family = * }

\raggedright
\usepackage{changepage}

\usepackage[aboveskip=1pt,labelfont=bf,labelsep=period,singlelinecheck=off]{caption}

\makeatletter
\renewcommand{\@biblabel}[1]{\quad#1.}
\makeatother

\usepackage{lastpage,fancyhdr,graphicx}
\usepackage{epstopdf}
\fancyheadoffset[L]{2.25in}
\fancyfootoffset[L]{2.25in}

\usepackage{color}

\definecolor{Gray}{gray}{.25}

\usepackage{graphicx}
\usepackage{float}
\usepackage{subfig}
\usepackage{sidecap}

\usepackage{wrapfig}
\usepackage[pscoord]{eso-pic}
\usepackage[fulladjust]{marginnote}
\reversemarginpar

\newtheorem{definition}{Definition}

\usepackage{amsmath}
\usepackage{mathtools}
\usepackage{amsfonts}
\usepackage[shortlabels]{enumitem} %enumerate abc

\usepackage{multirow}
\usepackage{adjustbox}

\newcommand\restr[2]{{% we make the whole thing an ordinary symbol
		\left.\kern-\nulldelimiterspace % automatically resize the bar with \right
		#1 % the function
		\vphantom{\big|} % pretend it's a little taller at normal size
		\right|_{#2} % this is the delimiter
}}

\newcommand{\sbrac}[1]{\left[ #1 \right]}
\newcommand{\cbrac}[1]{\left\{ #1 \right\}}
\newcommand{\eps}{\epsilon}
\renewcommand{\a}{\alpha}

\renewcommand{\a}{\alpha}

\renewcommand{\a}{\alpha}

\definecolor{darkgray}{rgb}{0.35, 0.35, 0.35}

\begin{document}
\vspace*{0.35in}

% title goes here:
\begin{flushleft}
{\Large
\textbf{A morphospace of functional configuration to assess configural breadth based on brain functional networks}
}
\newline
Duy Duong-Tran \textsuperscript{1}\textsuperscript{2},
Kausar Abbas \textsuperscript{1}\textsuperscript{2},
Enrico Amico \textsuperscript{1}\textsuperscript{2},
Bernat~Corominas-Murtra \textsuperscript{3},
Mario Dzemidzic \textsuperscript{4},
David Kareken \textsuperscript{4},
Mario Ventresca \textsuperscript{1}\textsuperscript{5} and
Joaqu\'{i}n Go\~{n}i \textsuperscript{1}\textsuperscript{2}\textsuperscript{6}*

\begin{singlespace}

\bigskip
\small
\ 1 School of Industrial Engineering, Purdue University, West-Lafayette, IN, USA
\\
\ 2 Purdue Institute for Integrative Neuroscience, Purdue University, West-Lafayette, IN, USA
\\
\ 3 Institute for Science and Technology, Am Campus 1, A-3400 Klosterneuburg, Austria
\\
\ 4 Department of Neurology, Indiana University School of Medicine, Indianapolis, IN, 46202 
\\
\ 5 Purdue Institute of Inflammation, Immunology and Infectious Disease, Purdue University, West Lafayette, IN, USA
\\
\ 6 Weldon School of Biomedical Engineering, Purdue University, West Lafayette, IN, USA
\\ 
\bigskip
* Corresponding email: \url{jgonicor@purdue.edu}
\end{singlespace}
\end{flushleft}
\justify
\begin{abstract}
\begin{singlespace}
\small 

The best approach to quantify human brain functional reconfigurations in response to varying cognitive demands remains an unresolved topic in network neuroscience.  We propose that such functional reconfigurations may be categorized into three different types: i) Network Configural Breadth, ii) Task-to-Task transitional reconfiguration, and iii) Within-Task reconfiguration. In order to quantify these reconfigurations, we propose a mesoscopic framework focused on functional networks (FNs) or communities. To do so, we introduce a 2D network morphospace that relies on two novel mesoscopic metrics,  Trapping Efficiency (\textbf{TE}) and Exit Entropy (\textbf{EE}), which capture topology and integration of information within and between a reference set of FNs. In this study, we use this framework to quantify the Network Configural Breadth across different tasks. We show that the metrics defining this morphospace can differentiate FNs, cognitive tasks and subjects.  We also show that network configural breadth significantly predicts behavioral measures, such as episodic memory, verbal episodic memory, fluid intelligence and general intelligence. In essence, we put forth a framework to explore the cognitive space in a comprehensive manner, for each individual separately, and at different levels of granularity. This tool that can also quantify the FN reconfigurations that result from the brain switching between mental states.

\end{singlespace}
\end{abstract}
%\newpage
\begin{section}*{Author Summary}
	\begin{singlespace}
	\small To understand and measure the ways in which human brain connectivity changes to accommodate a broad range of cognitive and behavioral goals, is an important question.  We put forth a framework that captures such changes by tracking  the topology and integration of information within and between FNs of the brain. Canonically, when FNs are characterized, they are separated from the rest of the brain network. The two metrics proposed in this work, Trapping Efficiency and Exit Entropy, quantify the topological and information integration characteristics of FNs while they are still embedded in the overall brain network. Trapping Efficiency measures the module’s ability to preserve an incoming signal from escaping its local topology, relative to its total exiting weights. Exit Entropy measures the module’s communication preferences with other modules/networks using information theory. When these two metrics are plotted in a 2D graph as a function of different brain states (i.e., cognitive/behavioral tasks), the resulting morphospace characterizes the extent of network reconfiguration between tasks (functional reconfiguration), and the change when moving from rest to an externally engaged “task-positive” state (functional preconfiguration), to collectively define network configural breadth.  We also show that these metrics are sensitive to subject, task, and functional network identities.  Overall, this method is a promising approach to quantify how human brains adapt to a range of tasks, and potentially to help improve precision clinical neuroscience.
%	Imagine the total repertoire of cognition as an abstract space where one can only take coarse snapshots of its complex topology; such space topology is completely unknown, otherwise. However, one wishes to determine the "shape" of such a space through what is available (still coarse snapshots). In the context of brain connectomics, we aimed to define and study this space through fMRI task and rest configurations. Functional connectomes can be viewed as sampled proxies (coarse snapshots) to represent, in part, the cognitive space. In this paper, we construct an example of a "cognitive space" using a mathematical mapping framework or a network morphospace. We define a minimally complex morphospace that captures two distinct features of mesoscopic structures in functional brain networks: i) Trapping Efficiency (\textbf{TE}) measuring the module ability to preserve the incoming signal from escaping its local topology, relative to its total exiting weights and ii) Exit Entropy (\textbf{EE}) measuring the module communication preferences with other communities  using information theory. Investigating the complex behavior of \textit{a priori} set of FNs, across different conditions, we show that modeling network configural breadth for a given subject's FN is equivalent to the study of neuromodulatory changes in FN across high/low cognitive demands imposed from tasks and rest. 
	
	\end{singlespace}
\end{section}

\section{Introduction}
Human behavior arises out of a complex interplay of functional dynamics between different brain networks \cite{bassett2011understanding}. These interactions are reflected in functional network reconfigurations as subjects perform different tasks or are at rest \cite{cole2014intrinsic,amico2019towards,amico2020disengaging}. One of the network neuroscience challenges is to develop a comprehensive framework to quantify the brain network (re-)configurations across different mental states and cognitive tasks. To that end, configurations across a collection of cognitive tasks can be conceptualized at three distinct levels of granularity:
\begin{itemize}
	\item \textbf{Network configural breadth} reflects the range of network functional configurations required to represent a given individual’s total repertoire of cognitive and emotional states. In practice, how well the entire “cognitive space” \cite{varona2016hierarchical,varoquaux2018atlases} is sampled depends on the number and choice of the tasks . This concept is inspired by Schultz et al. \cite{schultz2016higher}.
	\item \textbf{Task-to-task transitional reconfiguration} represents the specific shift in network functional configuration when a subject switches between cognitive/mental tasks \cite{douw2016state,gonzalez2015tracking}. For instance, task transitions and accompanying reconfigurations will occur when a subject transitions from quiet reflection to engage in a spatial problem solving task, or from a lexical retrieval to a decision making paradigm.
	\item \textbf{Within-task reconfiguration} represents specific network functional configuration changes that may occur within a single task. This phenomenon has been assessed  at the whole-brain level, showing the presence of distinct brain states within a task \cite{bassett2011dynamic,betzel2017positive,shine2016dynamics,shine2017principles,shine2019human}.
\end{itemize}

While brain network configural properties are task and subject dependent, \cite{schultz2016higher}, task-induced functional (re)configurations  are rather subtle in whole-brain functional connectomes, even when comparing task to rest \cite{cole2014intrinsic}. In addition, mesoscopic structures (e.g. functional networks of the brain) exhibit modular characteristics that adapt to cognitive demands without significantly affecting the rest of the system where higher levels of cognition emerge through the changing interactions of subsystems, instead of pairwise edge-level interactions \cite{bassett2011dynamic}. Hence, a mesoscopic scale (as the one provided by functional networks (FNs) or communities/modules) may uncover differential patterns of (re-)configuration, \cite{mohr2016integration}, across functional sub-circuits, which might otherwise not be detectable at other scales. Traditionally, a mesoscopic assessment of functional brain networks would involve the \textit{detection} of functional communities \cite{sporns2016modular} either based on topology (density-based) \cite{newman2006finding,newman2006modularity} or  on the information flow (flow-based) \cite{rosvall2008maps,rosvall2009map}.These approaches, however, are not designed to \textit{track} the dynamic behavior of \textit{a priori} set of communities across time, tasks, and/or subjects.
\begin{figure}[h!]
	\centering
	{\includegraphics[width=.8\columnwidth]{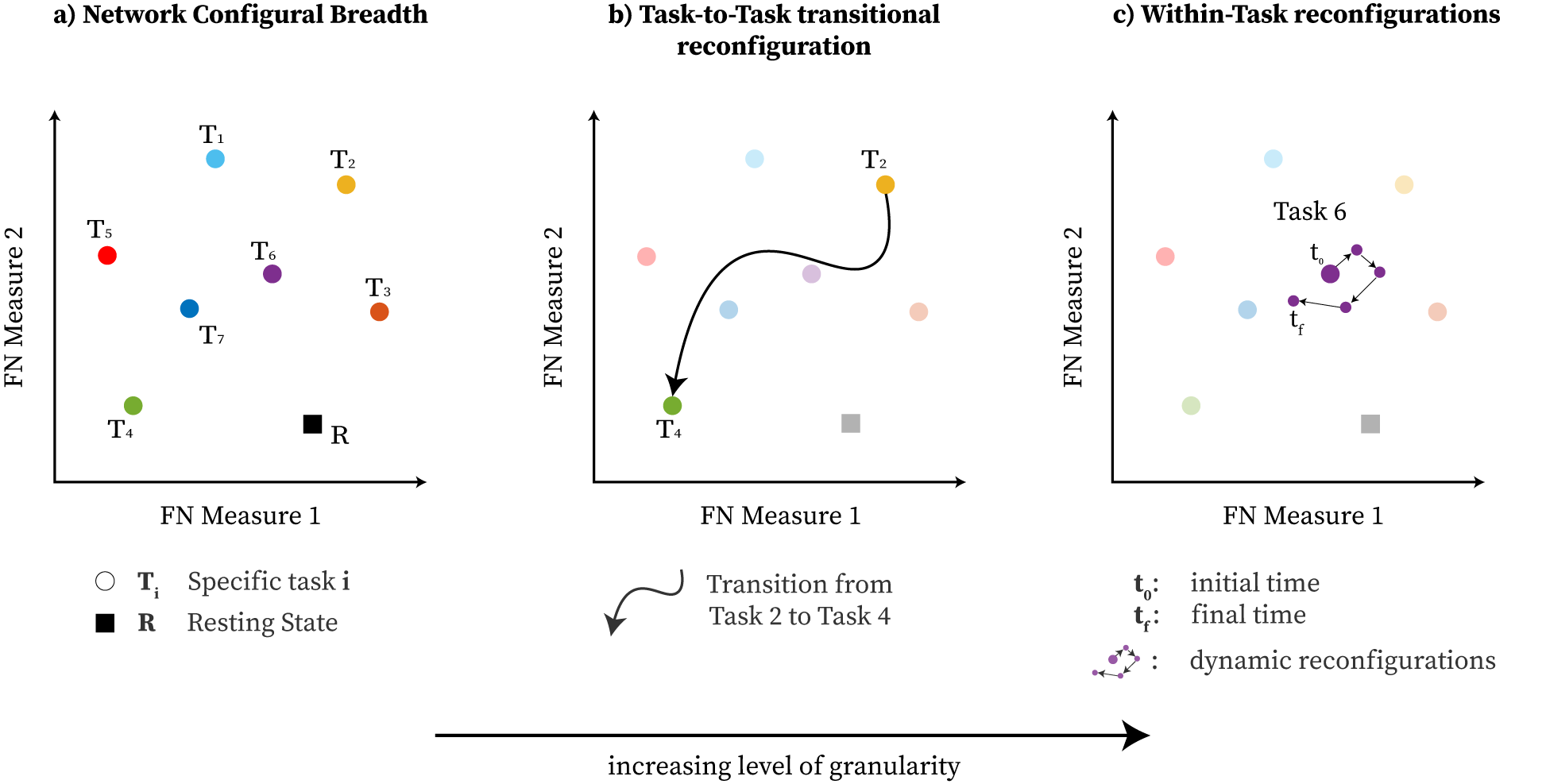} \label{fig:demon_configurations}}
	\caption{\small The three types of brain (re-)configurations that can be represented by a mathematical space parameterized by, in this case, two generic phenotypic measures of functional communities of the brain: \textbf{(a)} network configural breadth, which represents changes across a number of cognitive demands; \textbf{(b)} task-to-task transitional reconfiguration; and \textbf{(c)} within-task reconfiguration.} 
\end{figure}
The primary aim of this work is to clearly define and quantify different configurations that FNs can assume, as well as measure their nature of re-configurations switching between seemingly infinite number of cognitive states. From a graph-theoretical perspective, FNs and their corresponding reconfigurations are described by two attributes: topology and communication. From a system dynamic perspective, FNs can be characterized by segregation and integration \cite{sporns2013network} properties across which the human brain reconfigures across varied cognitive demands \cite{shine2016dynamics,shine2017principles,shine2018dynamic,shine2018low,shine2019human}. To formally capture these diverse characteristics of FNs, we constructed a mathematically well-defined and well-behaved 2D "mesoscopic morphospace" based on two novel measures defined for non-negative, undirected, weighted functional connectomes: Trapping Efficiency (\textbf{TE}) and Exit Entropy (\textbf{EE}). Trapping Efficiency captures the level of segregation/integration of a functional network embedded in the rest of the functional connectome and quantifies the extent to which a particular FN "traps" an incoming signal. Exit Entropy captures the specificity of integration of an FN with the rest of the functional connectome, and quantifies the uncertainty as to where (in terms of exit nodes) that same signal would exit the FN. In summary, this mesoscopic morphospace is a representation of the cognitive space as explored within and between cognitive states, as reflected by brain activity in fMRI. Such representation relies on FNs reconfigurations that can be tracked, at an individual level, and at different granularity levels in network (re-)configurations.

By using this 2D \textbf{TE,EE}-based morphospace, we formally study Network Configural Breadth (Figure \textbf{1(a)}), the most global and coarse grain exploration of the cognitive space,  and its subsequent functional configuration components. To that end, we formally define measures of (1) functional reconfiguration (capacity of an individual to reconfigure networks across widely differing cognitive operations) and (2) functional preconfiguration (efficiency of transition from resting-state to task-positive state \cite{schultz2016higher}, for potentially any community or FN. These measures are quantified for resting-state networks \cite{yeo2011organization} on the 100 unrelated subjects from the Human Connectome Project (HCP) dataset. We then study how such quantification is related to measures of cognitive abilities such as, fluid intelligence.

\section{A mesoscopic morphospace of functional reconfigurations}
The \textit{mesoscopic morphospace} proposed here is a two dimensional space built upon Trapping Efficiency and Exit Entropy measures for assessing functional networks or communities of functional connectomes. In this framework, functional connectomes must be undirected (symmetrical) weighted graphs, with \textit{non-negative} functional couplings. This framework allows for any \textit{a-priori} partition into functional communities. In this work, we assess the resting-state functional networks as proposed by Yeo et al. \cite{yeo2011organization} as the a-priori FNs. Also, we use functional connectivity (without incorporating structural connectivity information), which is a quantification of statistical dependencies between BOLD time-series of brain regions, and it can be used as a proxy of communication dynamics in the brain \cite{fornito2016fundamentals}. Under this section, further technical details that are not mentioned in the main text will be directed to specific section(s) in the Supplemental Information (\textbf{SI}).
\subsection{Computing mechanistic components for morphospace measures}
Given a functional network (community or module) in a whole-brain functional connectome (denoted as $\mathcal{C}\subset G(V,E)$), a mesoscopic morphospace is constructed to assess functional network behaviors through two focal lenses: level of segregation/integration (using graph topology), and specificity of integration (using information theory). We first define all necessary components to compute \textbf{TE} and \textbf{EE}. Specifically, for each functional community $\mathcal{C}(V_\mathcal{C},E_\mathcal{C})$, we
\begin{enumerate}[(i)]
	\item Define the set of states (denoted as $S$) for a given functional community which contains the set of transient states (denoted as $S_{trans}= \cbrac{V_{\mathcal{C}}}$), and absorbing states (denoted as $S_{abs}=\cbrac{j\mid w_{ij}>0; j\notin V_\mathcal{C},\forall i\in V_\mathcal{C}}$) such that $S=S_{trans}\cup S_{abs}$;
	\item Extract the induced (weighted) adjacency structure based on $S_{trans}$ and $S_{abs}$:
	\[
	\mathbf{A}\in (0,1)^{|S|\times |S|}
	\] 
	\item Construct the Terminating Markov Chain based on $S$ and $\mathbf{A}$; it is represented by
	$$\mathbf{P}\in  (0,1)^{|S|\times |S|}$$
	\item Compute the fundamental matrix (denoted as $\mathbf{Z}$) \cite{kemeny1960finite} which contains the mean number of steps a specific transient state in $S_{trans}$ is visited, for any pair of transient states in $S_{trans}$: 
	$$\mathbf{Z} = (\mathbf{I}_{|S_{trans}|} - \mathbf{P})^{-1} \in \mathbb{R}^{|S_{trans}|\times |S_{trans}|}_{+}$$ 
	where $\mathbf{I}_{|S_{trans}|}$ is the identity matrix of dimension $|S_{trans}|$;
	\item Compute the mean time to absorption (denoted as $\tau$) which contains the mean number of steps that the random particle needs to be absorbed by one of the states in $S_{abs}$, given that it states in some state in $S_{trans}$:
	\[
	\tau=\mathbf{Z}\mathbf{1}_{|S_{trans}|} \in \mathbb{R}^{|S_{trans}|\times 1}_{+}
	\] 
	where $\mathbf{1}_{|S_{trans}|}$ is the vector all ones of size $|S_{trans}|$;
	\item Compute the absorption probability matrix (denoted as $\Psi$), which contains the likelihood of being absorbed by one of the absorbing state, given that the stochastic process starts in some transient state:
	\[
	\Psi = \mathbf{Z}\sbrac{\restr{\mathbf{P}}{S_{trans},S_{abs}}} \in \mathbb{R}^{|S_{trans}|\times |S_{abs}|}_{+}
	\]
	where $\restr{\mathbf{P}}{S_{trans},S_{abs}}$ is the sub-transition probability matrix induced from (row) state $S_{trans}$ and (column) state $S_{abs}$.
\end{enumerate}

\subsection{Module Trapping Efficiency}
Module Trapping Efficiency, denoted as \textbf{TE} (unit: $\frac{steps}{weight}$), quantifies a module's capacity to contain a random particle from leaving its local topology, i.e. $\mathcal{C}$. Specifically, through FN topology, we want to assess its level of \textit{segregation/integration}, measured by the  $L_2$ norm of $\tau$ (unit: $steps$), i.e. the mean time to absorption of nodes in $\mathcal{C}$, normalized by its total exiting strength (unit: $weight$), measured by $\mathcal{L}_\mathcal{C}=\sum_{i\in S_{trans},j\in S_{abs}} A_{ij}$. Mathematically, trapping efficiency is quantified as follows:
\begin{align}
\mathbf{TE} &=\frac{||\tau||_2}{\mathcal{L}_\mathcal{C}} 
\end{align}
We see that the mean time to absorption vector, $\tau$, is dependent  on both \textbf{density-based} \cite{fortunato2010community,newman2006modularity} and \textbf{flow-based} \cite{rosvall2008maps,rosvall2009map,malliaros2013clustering} modularity. 
The numerator ($||\tau||_2$) measures the difference (through Euclidean $L_2$-norm) between the mean time to absorption of the functional module and its selected null module (the empty subgraph with the same number of nodes) in which $\tau_{null}=\vec{0}$. On the other hand, the denominator $\mathcal{L}_\mathcal{C}$ is a simple statistical summary of the module "leakage" to the rest of the cortex; hence, $L_1$-norm is chosen. The role of $\mathcal{L}_{\mathcal{C}}$ is to account for potential differences in trapping efficiency due to community size. Numerically, higher \textbf{TE} indicates that a module is more segregated (or equivalently, less integrated). This is because the FN topology traps the incoming signal efficiently, relatively to its exiting edges when embedded in the cortex. \textbf{TE} value ranges are given in \textbf{Fig. 2}.

\subsection{Module Exit Entropy}

Module Exit Entropy (denoted as \textbf{EE}, and in the range $\mathbf{EE}\in (0,1]$ and unitless) assesses the normalized level of uncertainty in selecting an exiting node in $S_{abs}$ of a random particle that starts in $\mathcal{C}$. The exit entropy, denoted as $\mathcal{H}_e$, measures the level of uncertainty exiting node $j\in S_{abs}$ (outside of the module) is preferred. Module exit entropy is mathematically formalized as:
\begin{equation}
\mathbf{EE} =\frac{\mathcal{H}_e}{\mathcal{N}_{\mathcal{C}}} = \frac{-\sum_{i=1}^{|S_{abs}|} \psi_i \log(\psi_i)}{\log (|S_{abs}|)}
\end{equation}
where preferential exit probability is the probability vector that contains $|S_{abs}|$ entries that represents the likelihood of an exit signal selects a specific exiting state $j\in S_{abs}$ such that $\sum_{j\in S_{abs}} \psi_j = 1$.

The numerator of $\mathbf{EE}(\mathcal{C})$, i.e. $-\sum_{i=1}^{|S_{abs}|} \psi_i log(\psi_i)$, measures the degree to which channels of communication between nodes in $S_{trans}$ and $S_{abs}$ are preferred for a fixed task/subject. It is noteworthy that $\mathbf{EE}$ is not influenced by the (cumulative) magnitudes (of functional connectivity values) that connect nodes from within the FN to outside (exiting) nodes. It is only affected by the distribution of such values. In particular, homogeneous distributions display high entropy levels and uneven distributions favoring certain exiting node(s) display low entropy. To demonstrate this point, an example is provided in \textbf{SI} under section \textbf{C.3}. The normalizer, $\mathcal{N}_{\mathcal{C}}=log(|S_{abs}|)$, is the maximum entropy obtained from a module in which all exit nodes have the same absorption rate. Numerically, a high \textbf{EE} would denote the homogeneous integration within the rest of the system whereas a low \textbf{EE} would indicate a preferential communication or integration of the module with the rest of the system. In terms of functional brain networks, module exit entropy facilitates the understanding of collective behavior from $\mathcal{C}$ to other FNs through its outreach channels (edges formed by nodes in $\mathcal{C}$ and exiting nodes in $G\setminus \mathcal{C}$). This is because entropy measures the level of uncertainty in communication; hence, lower entropy means higher specificity in communication between the FN with the rest of the cortex. \textbf{EE} value ranges are given in \textbf{Fig. 2}.
\subsection{The definition of the Mesoscopic Morphospace $\Omega$}
The two distinct features of each FN in brain graphs are addressed by a point $\mathbf{u}(\mathcal{C})$ in $\Omega \subset (0,M)\times [0,1]\subset \mathbb{R}^2$ as follows:
\begin{equation}
\mathbf{u}(\mathcal{C}) = (\mathbf{TE}(\mathcal{C}), \mathbf{EE}(\mathcal{C}))\in \Omega
\end{equation}
where $M< \infty$. for a given subject and task, a functional brain network $G$ is obtained with a pre-defined parcellation that results in $l$ induced subgraph $\mathcal{C}\subset G$, we can obtain $l$ points $\mathbf{u}(\mathcal{C})$ corresponding to $l$ FNs in network $G$.  
\begin{figure}[h]
	\centering
	{\includegraphics[width=.65\columnwidth]{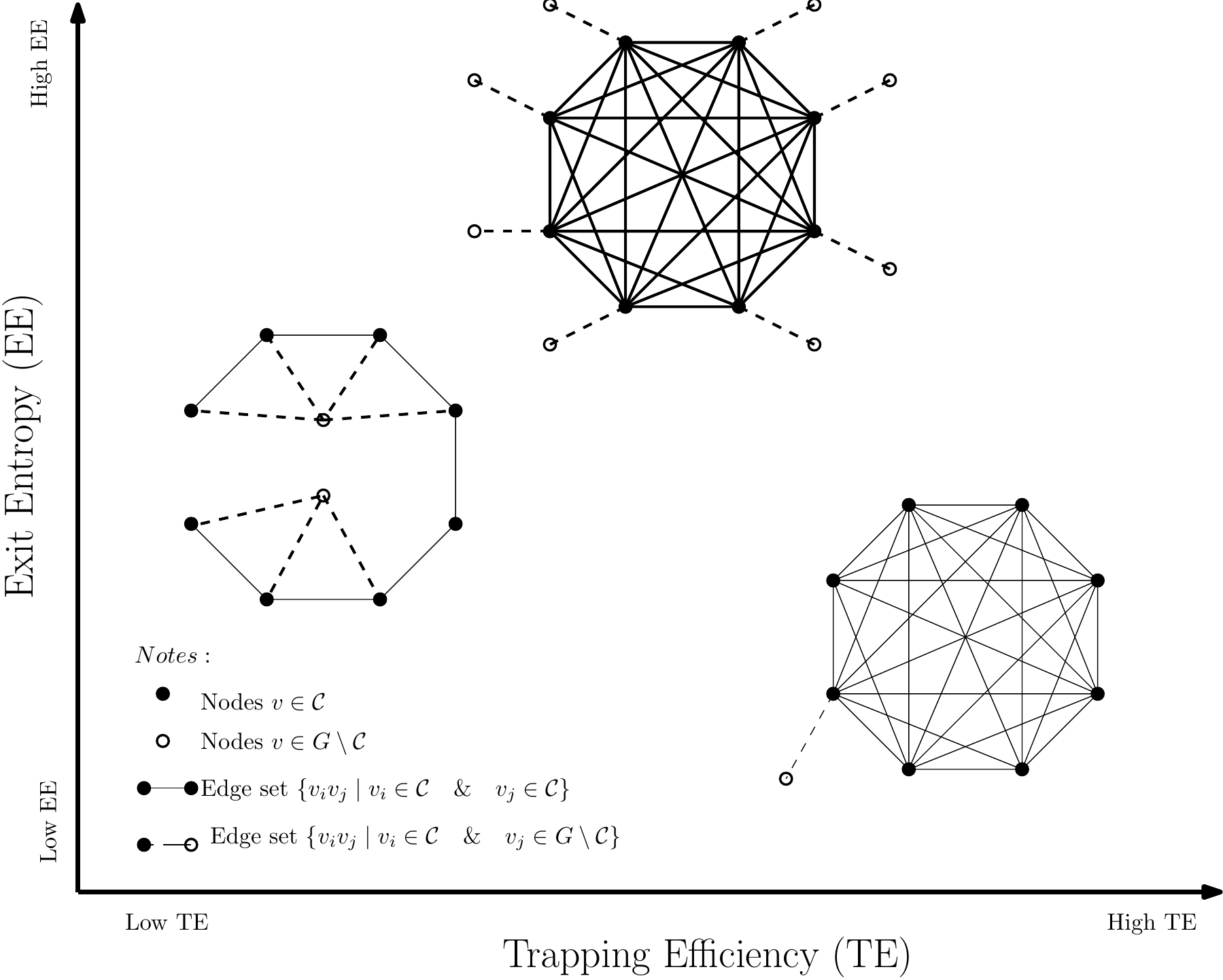} \label{fig:fig0}}
	\caption{\small \textbf{Morphospace Measurements - examples}: All three induced sub-graphs have the same cardinality ($|\mathcal{C}|=8$) with different number of exits (connections to $G\setminus \mathcal{C}$). Nonetheless, depending on their topological structures, the corresponding morphospace measurements (\textbf{TE} and \textbf{EE}) have rather distinct values.}
\end{figure}
In general, trapping efficiency, $\mathbf{TE}(\mathcal{C})$ is finitely bounded by construction (see more details in \textbf{Section D.6} under \textbf{SI}). However, a better bound is possible for the HCP dataset. This is due to two driving factors: connectome sparsity and edge weights \cite{avena2015network}. We address the upper bound for \textbf{TE} as: $\max (\mathbf{TE(\mathcal{C})})= M = 1$. In terms of $\mathbf{EE}(\mathcal{C})$, its numerical range $\mathbf{EE}(\mathcal{C})\in (0,1]$. Hence, $\Omega\subset (0,1)\times [0,1]$.

\section{The network configural breadth formalism}

Studying the manifold topology defined in this 2D mesoscopic morphospace theoretically requires an infinite amount of points. In finite domain with discrete sampling of the morphospace, polytope theory, a mathematical branch that studies object geometry, allows us to create a reasonable scaffold presentation with well-defined properties to formally define and quantify configural components of the functional networks.

Polytope theory is a branch of mathematics that studies the geometry of shapes in a $d-$dimensional Euclidean space, $\mathbb{R}^d$. Given a set of points in this space, $W=\cbrac{\mathbf{x}_1,\mathbf{x}_2,....\mathbf{x}_{|W|}}$, a convex hull formed by $W$ is represented by 
\[
\mathbf{Conv}(W)= \cbrac{\sum_{j=1}^{|W|} \a_j \mathbf{x}_j \mid \sum_{j=1}^{|W|} \a_j=1, \a_j\ge 0}
\]
One can compute the notion of volume of the convex hull enclosed by $\mathbf{Conv}(W)$, denoted as  $\mathit{Vol}(\mathbf{Conv}(W))$. Given that the morphospace is 2D, the manifold dimension can be from 0 up to 2. In the \textbf{SI} under Section \textbf{D.1.}, further details on volume computation are defined.
\newline

The functional network configural breadth, for the $i^{th}$ subject, is compartmentalized into two components: 
\begin{itemize}
\item FN (task) reconfiguration and 
\item FN rest-to-[task-positive] preconfiguration. 
\end{itemize}
We then propose a mathematical relation between network configural breadth with FN reconfiguration and preconfiguration as follows:
\begin{equation}
\mathcal{F}_{i} = f(\mathcal{R}^{FN}_{i}, \mathcal{P}^{FN}_{i})
\end{equation}
where $\mathcal{F}_{i}$ represents configural breadth for subject $i^{th}$. Here, we provide directly the measures that quantify (functional) reconfiguration and preconfiguration of FNs for $i^{th}$ subject's configural breadth. Tasks are assigned the same level of importance and hence, no task is weighted more than others. 
\begin{figure}[h]
	\centering
	{\includegraphics[width=.8\columnwidth]{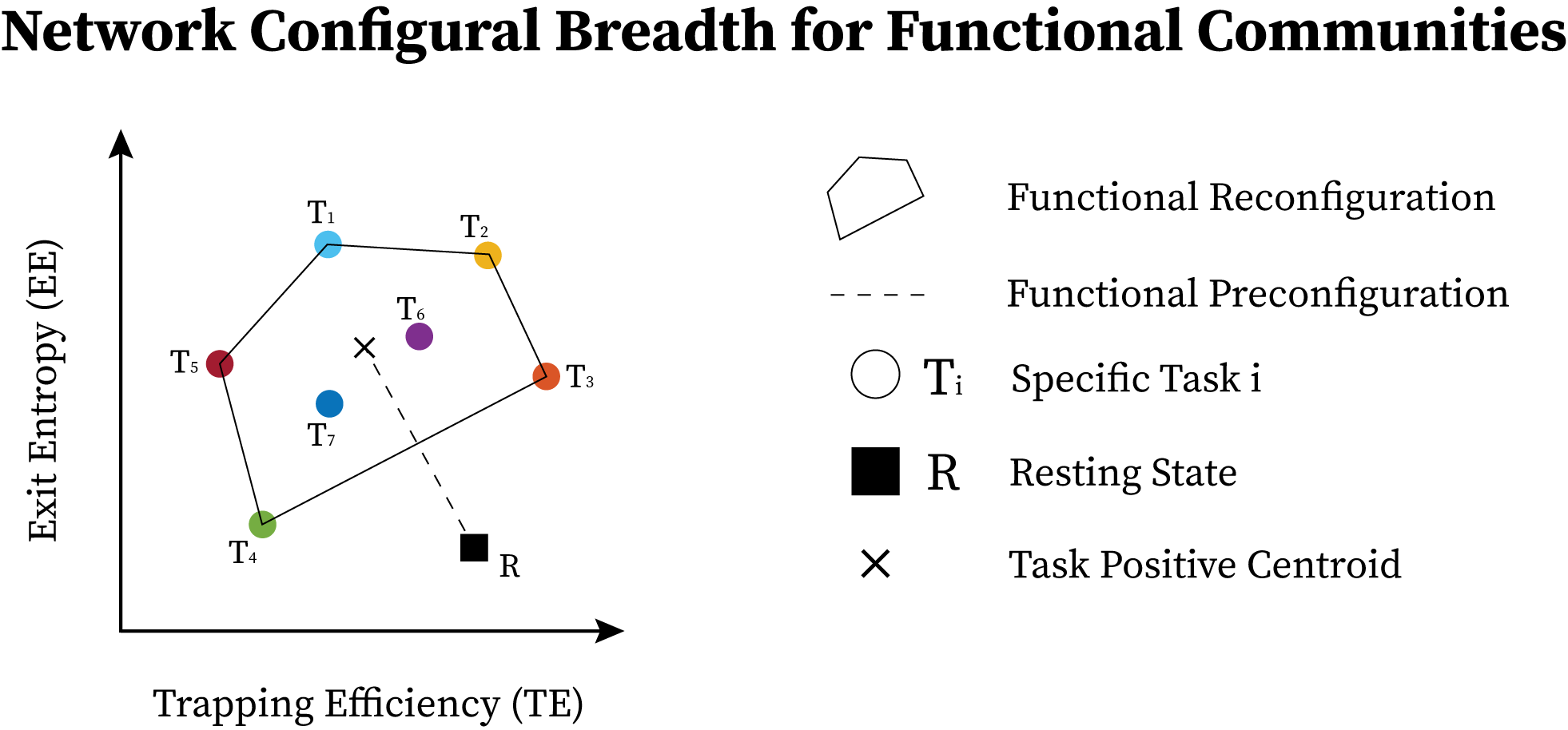} \label{fig_p/reconfigdemo}}
	\caption{\small \textbf{Functional Network configural breadth} is geometrically represented using two predefined morphospace measures. Specifically, for mesoscopic structures such as communities in functional brain networks, the first measure is Trapping Efficiency (\textbf{TE}) while the second is is Exit Entropy (\textbf{EE}). In this case, tasks T1 to T5 belong to the convex hull (e.g. Pareto front - further details are available in \textbf{SI} section \textbf{D}) while T6 and T7 is in the interior enclosed by the convex hull.}
\end{figure}
\subsection{Functional Reconfiguration}
\begin{definition}
Functional reconfiguration in this work is represented by a 2-dimensional spatial volume derived from given FN’s \textbf{EE} and \textbf{TE} coordinate values.  As such, it represents an example of “cognitive space” \cite{varona2016hierarchical,varoquaux2018atlases} within a functional domain that spans a variety of network states under various task-evoked conditions.  We  quantify this as
	\begin{equation}
	\mathcal{R}^{FN}_{i} = \mathit{Vol}(\mathbf{Conv}(W^{FN}_i))
	\end{equation}
	where $W^{FN}_i$ represents the set containing all investigated task coordinates of subject $i$'s FN; $\mathbf{Vol}(Conv(W^{FN}_i))$ is the convex hull volume induced by points in $W^{FN}_i$.  
\end{definition}
For a given subject $i^{th}$'s FN, note that $\mathbf{Conv}(W^{FN}_i$ represents the broad span (breadth) of task configurations for a given functional community. Subsequently, $\mathcal{R}^{FN}_{i}$ represents the amount of breadth as measured by the volume of $\mathbf{Conv}(W)$. Functional reconfiguration for a given subject's FN, denoted as $\mathcal{R}^{FN}_{i} $, is geometrically depicted in \textbf{Fig. 3}.
\subsection{Functional Preconfiguration}
\begin{definition}
Functional preconfiguration reflects the topologically distributed equipotentiality that is theoretically designed to enable an efficient switch from a resting state configuration to a task-positive state \cite{schultz2016higher}, and is quantified as follows
	\begin{equation}
	\mathcal{P}^{FN}_i = ||Rest^{FN}_i - \eta_{W^{FN}_i}||_2
	\end{equation}
	where $\eta_{W^{FN}_i}$ is the geometrical centroid of $W^{FN}_i$; $\mathcal{P}$ measures the distance between rest to task-general position (represented by $\eta_{W^{FN}_i}$). It is defined with the selected metric space, in this case is the $2$-norm in Euclidean space. 
\end{definition}

Note that functional preconfiguration can be viewed as $ \mathit{Vol}(\mathbf{Conv}(W))$ where the convex hull is defined solely by two points: FN's rest and FN's geometrical centroid of task convex hull, i.e. $W=\cbrac{R^{FN}_i,\eta_{W^{FN}_i}}$. In such regards, the notion of $ \mathit{Vol}(\mathbf{Conv}(W))$ is also suitable to describe the configural breadth between rest and task positive location. Functional preconfiguration is geometrically depicted in \textbf{Fig. 3}.

%\begin{remark}
%Given the concept of functional reconfiguration and preconfiguration, second order and first order measurements, respectively, are the most suitable. Note that one cannot directly compare the numerical value of functional preconfiguration with corresponding reconfiguration because of the difference in computational approach. Nonetheless, one can directly compare the numerical values within FN's pre- or re- configurations. 
%\end{remark}

\section{Results}
The mesoscopic morphospace formalized in section \textbf{2} is used to assess network configural breadth in terms of functional preconfiguration and reconfiguration for the one hundred unrelated subjects (HCP, Q3 release). This dataset includes (test and retest) sessions for resting state and seven fMRI tasks: gambling (GAM), relational (REL), social (SOC), working memory (WM), language processing (LANG), emotion (EMOT), and motor (MOT). Whole-brain functional connectomes estimated from this fMRI dataset include 360 cortical brain regions \cite{glasser2016multi} and 14 subcortical regions. The functional communities evaluated in the morphospace include seven cortical resting state FNs from \cite{yeo2011organization}: visual (VIS), somatomotor (SM), dorsal attention (DA), ventral attention (VA), frontoparietal (FP), limbic (LIM), default mode (DMN) and one comprised of subcortical regions (SUBC). Additional details about the dataset are available in \textbf{SI}.

\subsection{Task- and subject-sensitivity} 
	\subsubsection{	Within- and between-subject task sensitivity}
We first evaluate the capacity of module trapping efficiency and exit entropy to differentiate between tasks \textbf{within} subject \textbf{Fig. 4A}. For both test and retest sessions of each subject, we compute the \textbf{TE} and \textbf{EE} metrics for each FN. We compute these values for all 8 fMRI conditions. We compute the intraclass correlation coefficient (ICC), with test and retest (per subject) being the repeated measurements and task being the class variable (\textbf{TE} in \textbf{Fig. 4A-top} and \textbf{EE} in \textbf{Fig. 4A-bottom}, respectively, where each ICC is computed using a 2 (test, retest,) by 7(tasks) design, and the ICC reflects task within-subject sensitivity). For most subjects, ICC values in all FNs are high and positive values. \textbf{EE} displays a higher within-subject task sensitivity than \textbf{TE}. Specicifically, \textbf{TE} in VIS, DA and DMN most distinguished between the cognitive tasks, whereas \textbf{EE} in VA and FP was best at distinguishing the within-subject task-based configural changes. The ICC values for both coordinates were the lowest for LIM. 
\begin{figure}[H]
	\centering
	{\includegraphics[width=\columnwidth]{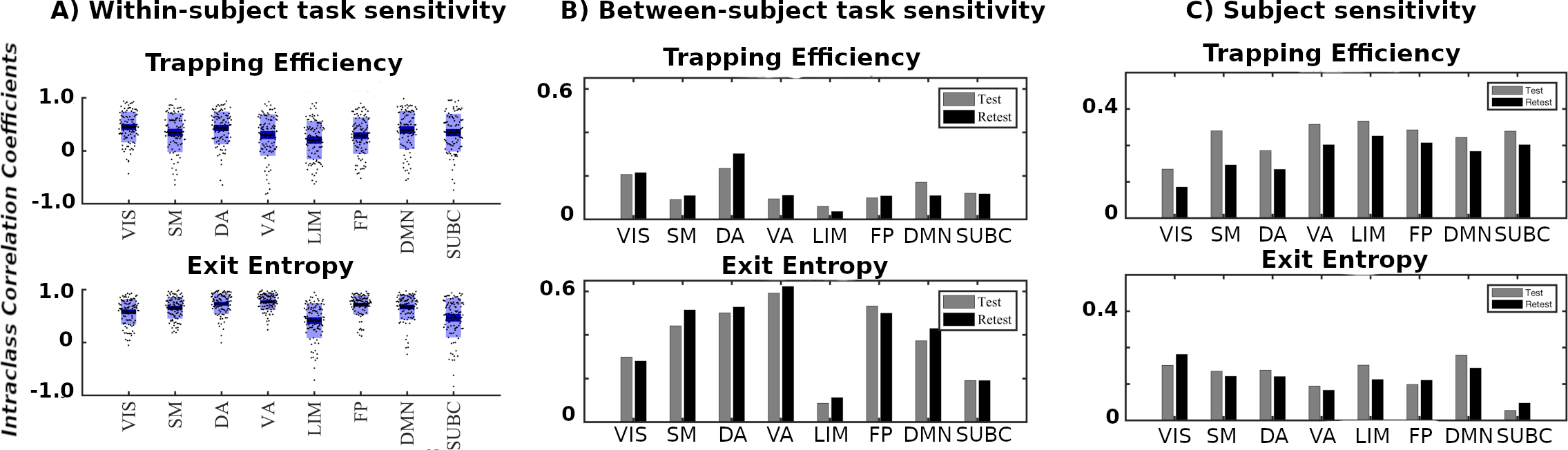}\label{fig:M.Es.E task ICC}}
	\caption{\small \textbf{Morphospace measures and their task- and subject- sensitivity measured by intra-class correlation coefficients for each functional network}: \textbf{(A)}Within-subject task sensitivity of Module trapping efficiency (\textbf{TE}) and exit entropy (\textbf{EE}) for each FN per subject . \textbf{(B)} Between-subject task sensitivity of \textbf{TE} (top) and \textbf{EE} (bottom). \textbf{C)} Subject-sensitivity ICC of \textbf{TE} (top) and \textbf{EE} (bottom).}
\end{figure}
We then evaluate the degree to which morphospace metrics capture cohort-level configural changes. To test this, for each morphospace metrics (\textbf{TE} or \textbf{EE}), we compute ICC of each FN where subjects as the repeated measures and task the class variable (\textbf{Fig. 4B}). We performed the evaluation separately for test and retest sessions as denoted by gray and dark bars, respectively, for \textbf{TE} (\textbf{Fig. 4B-top}) and \textbf{EE} (\textbf{Fig. 4B-bottom}). \textbf{EE} captures cohort-level task-based signatures as ICC values are consistently higher than those of \textbf{TE}. Interestingly, LIM has the lowest cohort-level task-based sensitivity for both morphospace metrics.    
\subsubsection{ Subject sensitivity across tasks}
Here, we compute ICC considering the tasks (fMRI conditions) the repeated measurements and considering subjects the class variable (\textbf{Fig. 4C}). It is note-worthy that \textbf{TE} is superior in uncovering subject fingerprints, compared to \textbf{EE}, for the majority of FNs.  This is complementary to \textbf{EE} being more task-sensitive.
\subsubsection{\textbf{TE} and \textbf{EE} are disjoint features}
Results in the sections \textbf{4.1.1. and 4.1.2.} suggests that \textbf{TE} and \textbf{EE} have the differentiating capacity to highlight non-overlapping characteristics of objects under consideration, i.e. task- and subject- based FNs. First of all, for within-subject task differentiation (\textbf{Fig. 4A}), FNs with high ICC values in one measure do not necessarily show a similar tendency in the other. For instance, VA has the third lowest mean \textbf{TE} value in characterizing within-subject tasks differentiation but it has the highest mean \textbf{EE} score. Similarly, FP has second lowest average \textbf{TE} score while third highest \textbf{EE} score indicating that each of the two measure caputures unique aspects of a given FN. Secondly, evidence of disjoint features is shown through the ICC results in cohort-level task-sensitivity (\textbf{Fig. 4B}) and subject-sensitivity (\textbf{Fig. 4C}) configural changes. Indeed, \textbf{TE} is superior in detecting subject fingerprints while \textbf{EE} is better in unraveling task fingerprints. The idea is that, for a given studied object (i.e. task-based FNs), configurations are shown to “stretch” in exclusive/disjoint directions (subject-sensitive trapping efficiency and task-sensitive exit entropy).

\subsection{Quantifying network configural breadth on functional networks}
The mesoscopic morphospace allows the quantification of network configural breadth. For a given functional community, we compute functional reconfiguration (degree of configurations across tasks) and preconfiguration (distance from rest to task positive state), using formula \textbf{(5,6)}, respectively. 

\subsubsection{Group-Average Results}
The group average behavior of functional communities is shown in \textbf{Fig. 5}. Functional reconfiguration of FNs are shown as filled convex hulls whereas preconfiguration of FNs are shown as dashed lines from rest to the corresponding task hull geometric centroid. 
\begin{figure}[H]
	\centering
	{\includegraphics[width=.85\columnwidth]{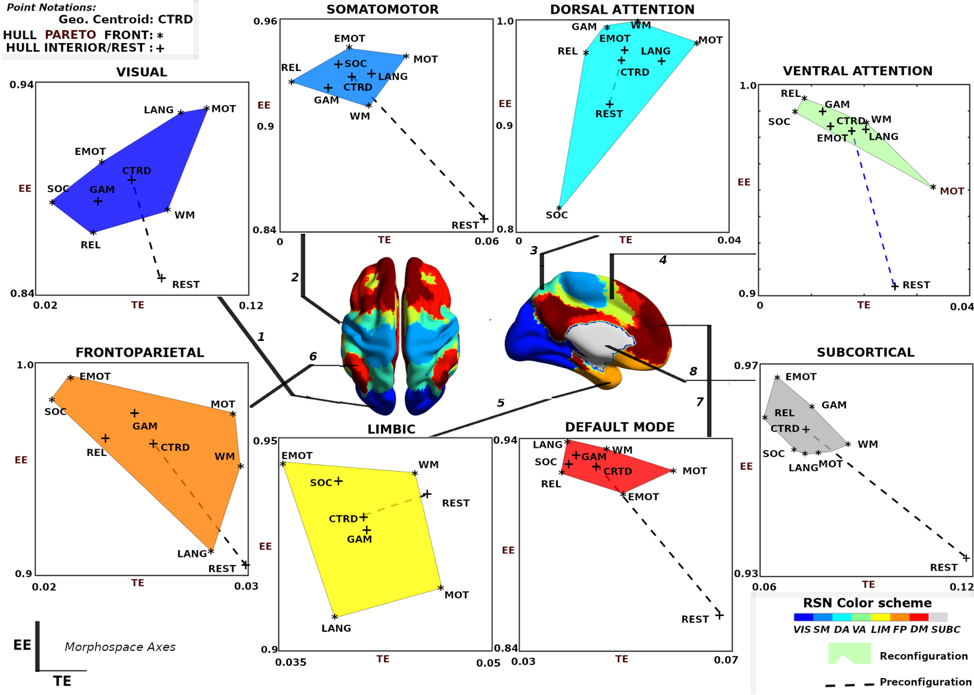} \label{fig:geo_presentation_FF}}
	\caption{\small \textbf{Visualization of Network Configural breadth}: functional reconfiguration and preconfiguration for all FNs are represented using group average of individual subjects' coordinates. Task coordinates in this space are represented by either asterisk (*) or the plus (+) symbol. The asterisk symbol is used for those tasks that are part of the Pareto front of the convex hull; the plus symbol represents either the resting state or task that belongs to the interior of the convex hull. Note that $x-$ and $y-$ axis are purposely not scaled in the same range so that the full range of values for all tasks, task-centroid, and rest can be more easily visualized.}
\end{figure}
To facilitate comparing network configural breadth across all functional networks, these same convex hulls are shown in \textbf{Fig. 6A} with the same $x-$ and $y-$ axis values. VIS network polytope, representing group-average behavior, is lower in \textbf{EE} relative to other FNs. 
With the exception of VIS and SUBC, all other FNs cluster in a similar, high \textbf{EE}/ low \textbf{TE} area of the morphospace (\textbf{Fig. 6A}). It should be noted that different tasks and subject populations (e.g., older or clinical groups) might cluster FNs differently. We also note that the subcortical polytope is relatively high in exit entropy. However, the subcortical parcellation might not optimally reflect the functional and/or structural makeup of various subcortical regions (e.g., role of the basal ganglia in the motor system) so these results should be interpreted cautiously.

One observation drawn from such a presentation is that the morphospace framework reconfirms, quantitatively, that functional dichotomy of the brain between task-positive and rest state \cite{fox2005human}. Specifically, the default mode network acts more as a segregated module with high level of integration specificity at rest - as seen in the lower right regime with high \textbf{TE}, low \textbf{EE} values - as opposed to under task-evoked conditions - as seen in the top left corner with low \textbf{TE}, high \textbf{EE} values (\textbf{Fig. 5} \textit{Default Mode}) \cite{greicius2003functional,fox2005human}.

Another observation is that in terms of segregation level measured by \textbf{TE}, the lower bound of subcortical convex hull is, approximately, the upper bound of other FNs, with the exception of the visual network. Figures \textbf{7.1A} and \textbf{7.2A} also summarize functional reconfiguration and preconfiguration respectively, for test and retest fMRI sessions in all subjects and FNs. Here, the VIS system displays the largest functional reconfiguration (see \textbf{Fig. 7.1A}). From \textbf{Fig 7.2A}, functional preconfigurations display a more comparable magnitude among all FNs.
\begin{figure}[H]
	\centering
	{\includegraphics[width=\columnwidth]{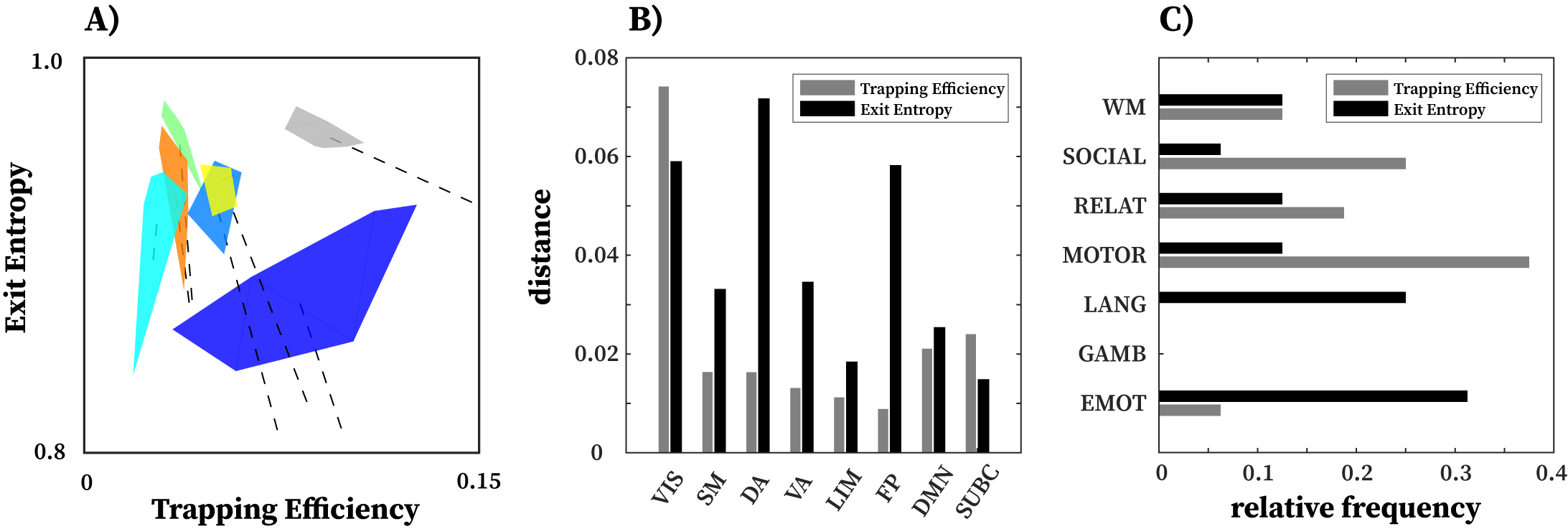} \label{fig:overlaid_geometry}}
	\caption{\small \textbf{(A)} An illustration of network configural breadth for all functional communities. \textit{Polytope colors are analogous to the ones scheme shown in \textbf{Fig. 5}}. For each functional community, the dash line represents the amount of functional preconfiguration whereas the polytope volume represents the amount of functional reconfiguration. \textbf{(B)} Maximal distance is computed using the maximum pairwise distance between two tasks for a given functional network. \textbf{(C)} Relative frequency with which a task appears in the maximal distance normalized by 16 (8 FNs and 2 task per FN).}
\end{figure}
Further evidence of disjoint feature is also displayed in \textbf{Fig. 6B} and \textbf{6C}. In \textbf{Fig. 6B}, maximal distance is computed using pairwise distances for two given tasks for a specific FN. The result shows that for a given FN, the two measures complement each other and in many cases, stretch the cognitive space in one direction or the other. For instance, in case of DA and FP, the maximal distance in \textbf{EE} is very high but low for \textbf{TE} whereas in VIS and SUBC, \textbf{TE} maximal distance is higher than that of \textbf{EE}. Furthermore, in \textbf{Fig. 6C}, only specific tasks (e.g. Motor and Emotion) that push the cognitive space in particular direction (which is captured by maximal distance computation). Evidence of disjoint features is also illustrated by the relative frequency of Motor and Emotion tasks for which \textbf{TE} and \textbf{EE} are complimentary.

\subsubsection{Subject specificity of pre- and reconfiguration of functional networks}
The formulation of network configural breadth (in terms of preconfiguration and reconfiguration) enables us to assess these properties at the subject level. %To what extent is network configural breadth a fingerprint of subjects and to what extent we can detect this with such framework? 
In \textbf{Fig 7.1B and 7.2B}, we use ICC to analyze the ability of morphospace measures (in the form or reconfiguration (panels \textbf{Fig. 7.1}) and preconfiguration (panels \textbf{Fig. 7.2})) to reflect subject identity within each FN. For all FNs from Yeo et al. \cite{yeo2011organization}, the ICCs suggest that subjects can be differentiated from each other when contrasted against a corresponding null model (see \textbf{SI} Section \textbf{D.4.1} for details). We see that subject sensitivity scores of all eight FNs for both pre- and re- configurations are higher than their corresponding null models. Finally, for a fixed FN, functional preconfigurations dominated the subject sensitivity ranking, as illustrated by \textbf{Fig. 7C}. Furthermore, FP, DMN and VA preconfigurations are among the FNs with highest subject fingerprints in overall subject sensitivity ranking. 
\begin{figure}[H]
	\centering
	{\includegraphics[width=\columnwidth]{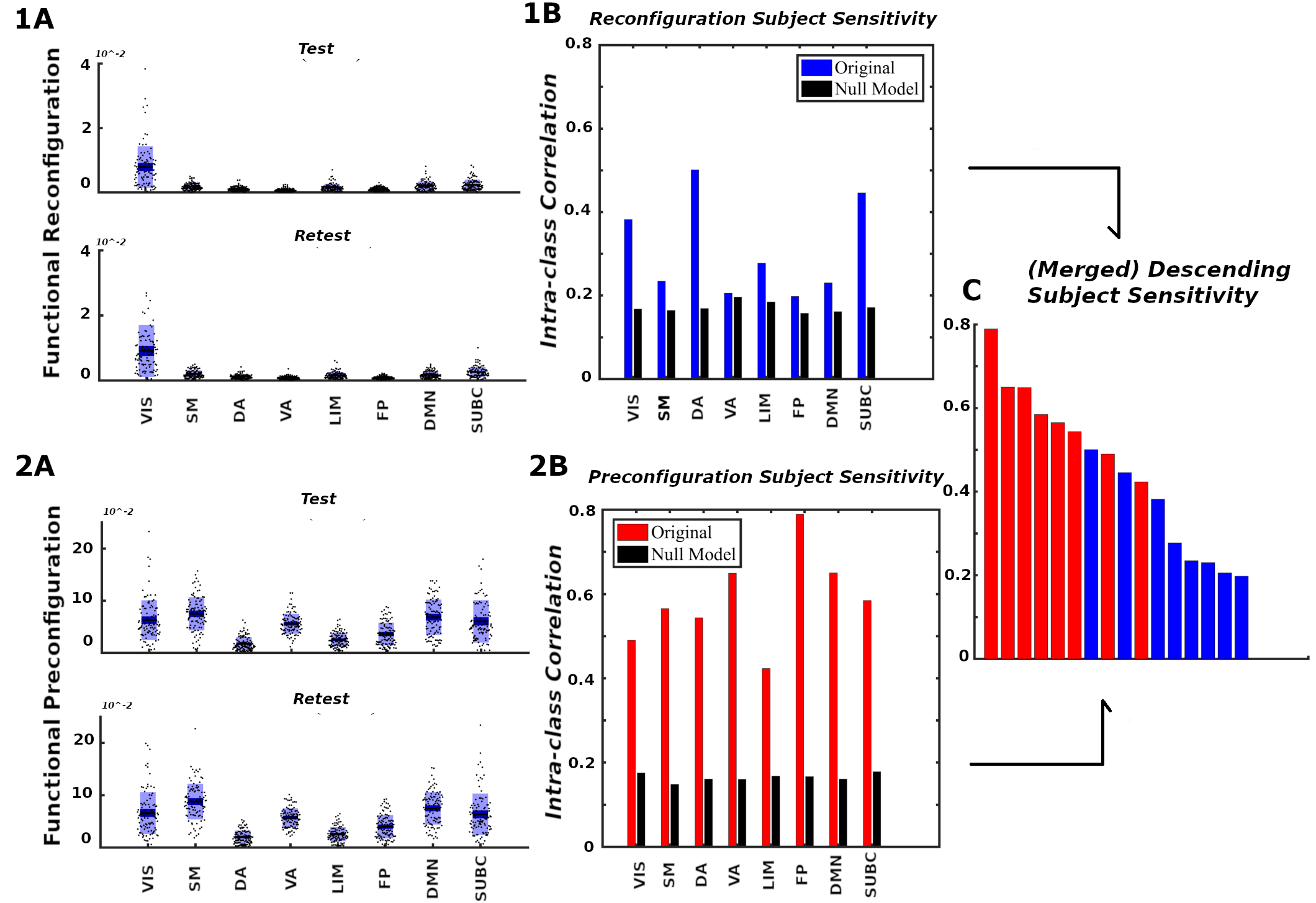} \label{fig:fig2_maintext_func_flex}}
	\caption{\small \textbf{Network Configural breadth - subject specificity analysis}: Panels \textbf{1} and \textbf{2}show functional reconfiguration and preconfiguration, respectively, from both magnitude and subject-sensitivity viewpoints. For each functional network, the \textbf{A} Panels report subject's preconfiguration and reconfiguration values whereas the \textbf{B} Panels quantify subject sensitivity. Reconfiguration and preconfiguration measures are displayed in blue and red, respectively. Panel \textbf{C} merges all 16 configural breadth terms in descending order of subject sensitivity. Color version of this figure is available online.}
\end{figure}
\subsection{Network configural breadth and behavior}
Network configural breadth, compartmentalized into FN reconfiguration $\mathcal{R}^{FN}$ and preconfiguration $\mathcal{P}^{FN}$, shows high level of subject sensitivity. This allows us to assume that $\mathcal{F}_i$ is associated with an individual's behavioral measures (denoted as $\mathbb{m}_i $ for subject $i^{th}$). Serveral studies reported that FP and DMN networks are associated with memory and intelligence \cite{gray2003neural,schultz2016higher,tschentscher2017fluid}. Therefore, we evaluated if the outlined framework reflects  four widely studied cognitive/behavioral measures, related to memory and intelligence: episodic memory, verbal episodic memory (Verb. Epi. Mem.), fluid intelligence $gF$, and general intelligence $g$. While fluid intelligence reflects subject capacity to solve novel problems, general intelligence, $g$, reflects not only fluid intelligence, $gF$, traits but also crystallized (i.e. acquired) knowledge (\cite{cattell1963theory} and typically denoted as $gC$). The early notion of general intelligence is conceptualized by Spearman’s positive manifold \cite{spearman1904general} that cannot be fully described using a single task. Quantification of g can be accomplished using subspace extraction techniques such as explanatory factor analysis (\cite{dubois2018distributed}) or principal component analysis (PCA \cite{schultz2016higher}). In this work, we quantified $g$ using the PCA approach described in Schultz and Cole \cite{schultz2016higher}.Mathematically, we propose the following composite relationship:
\begin{align}
	\mathbb{m}_i &= \Upsilon (\mathcal{R}^{FN}_{i}, \mathcal{P}^{FN}_{i})
\end{align}
\begin{figure}[H]
	\centering
	\includegraphics[width=.9\columnwidth]{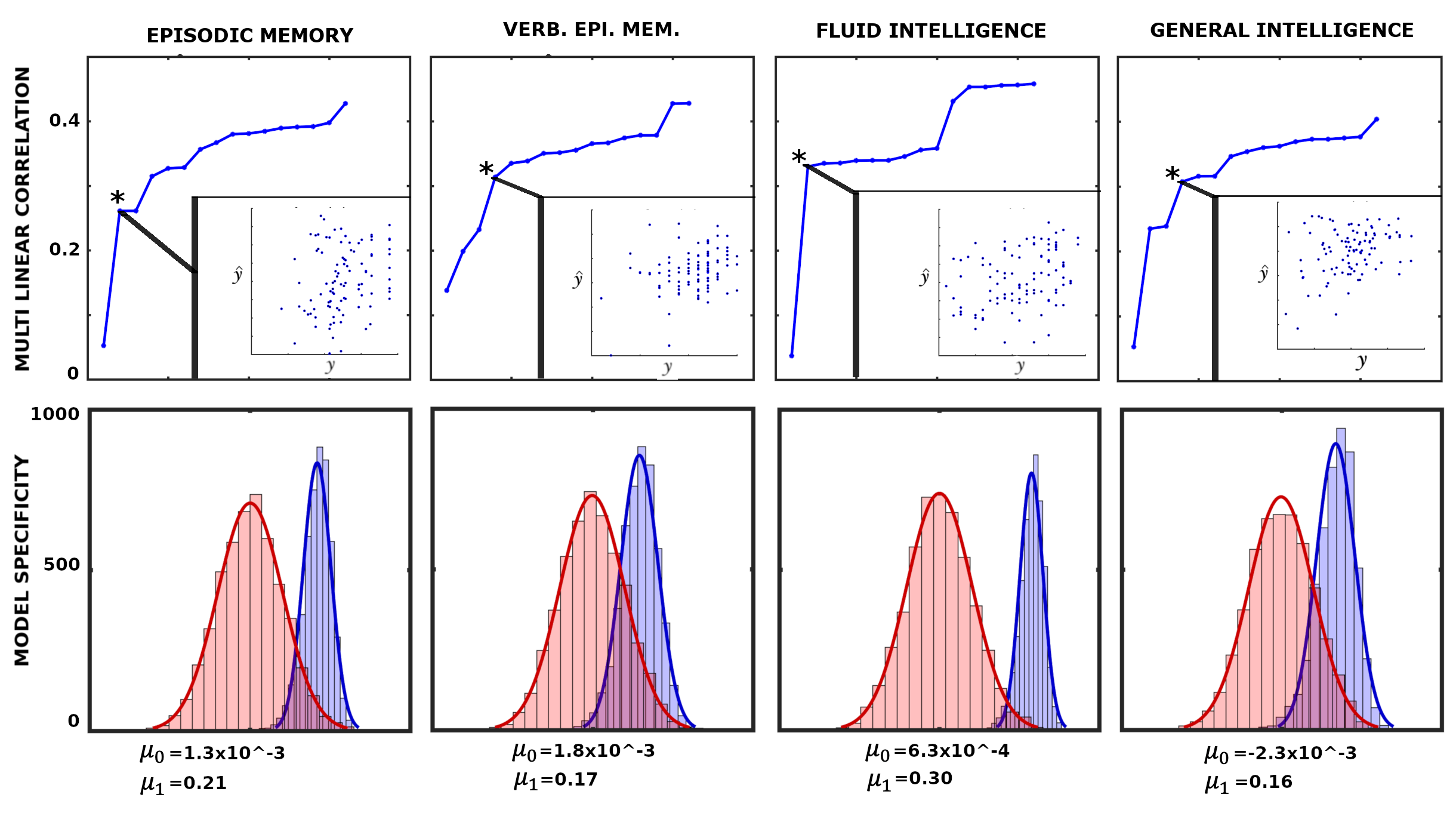}
	\label{fig:Func_CorrelationAnalysis}	
	\caption{\small \textbf{Associations between network configural breadth and behavior}:  the $x$-axis represents functional network preconfiguration and reconfiguration terms, i.e.  $\mathcal{P}^{FN}_i$ and $\mathcal{R}^{FN}_i$, ordered in decreasing subject fingerprints (as shown in \textbf{Fig. 7C}). The \textit{top panels} illustrate iterative multilinear regression model (MLM) while \textit{bottom panels} show model specificity (MS) for corresponding behavioral measures. Asterisk represents the optimal MLM with lowest p-value. Further details available in \textbf{SI} section \textbf{E.1.}}
\end{figure}
Having established a plausible connection between behavioral measures and $\mathcal{P}^{FN}$, $\mathcal{R}^{FN}$, equation \textbf{(7)} can be viewed as a multi-linear model (MLM) using FN preconfiguration and reconfiguration as independent variables (or predictors). The MLM is constructed iteratively, starting with the descriptor with the highest individual fingerprints in \textbf{Fig. 7C}. In each iteration, the subsequently ranked descriptor (according to \textbf{Fig. 7C}) is appended to the existing ones. The best MLM (denoted with an asterisk in \textbf{Fig. 8}), which determines the number of linear descriptors included the model, is selected based on the model p-value.

%Constructing the MLM to infer the intrinsic relationship is a necessary but not sufficient if the ultimate goal is to discover if there is a truly robust relationship between network configural breadth and behavioral measures. If there exists such robustness, then there has to be a certain degree of specificity in these models such that only significant correlations are observed when linear predictors are correlated with the true behavioral measures. Specifically, network configural breadth, as mathematically formulated using linear descriptors, must show that it is strongly correlated with a designated measures and not anything else, say a randomized vector.

To test the level of specificity in the model, we performed 2000 simulations of $k-$ fold cross validation where $k=5$ between the selected MLM and the corresponding behavioral measure. Specifically, for each cross validation (per simulation), we obtain a correlation between the 20 left-out values (y) with the predicted values ($\hat{y}$). Hence, in each simulation we obtained five correlation and their mean value. It can be shown that those means follows a normal distribution (details shown in SI). Lastly, to provide the level of specificity of linear descriptors, we present a corresponding null model where the same descriptors are evaluated to predict random vectors of appropriate size. To test our model and its ability to predict the behavioral measures, we rely completely on network configural breadth predictors ranked in descending order of subject specificity.

\begin{table}[h]
	\centering
	\begin{adjustbox}{max width=\columnwidth}
		\begin{tabular}{|c|c|c|c|c|c|}
			\hline
			\multirow{2}{8.5em}{\bfseries MLM terms \\/coefficients} & Constant & $\mathcal{P}^{FP}$ & $\mathcal{P}^{DMN}$ & $\mathcal{P}^{DA}$ & $\mathcal{P}^{SUBC}$ \\\cline{2-6}
			& \bfseries	$\beta_0$ & \bfseries	$\beta_1$& \bfseries	$\beta_2$& \bfseries	$\beta_3$& \bfseries	$\beta_4$\\
			\hline\hline
			\bfseries Episodic Memory & 0.6 & 2.9 & -9.3 &  &\\
			\hline
			\bfseries Verbal Episodic Memory & 0.5 & 11.8 & -1.1 & -8.8 & -6.1\\			
			\hline
			\bfseries $gF$ & 0.7 & 5.1 & -12 & & \\
			\hline
			$g$ & 0.8 & 3.9 & -5.5 & -3.6 & -5.7 \\ 
			\hline\hline
		\end{tabular}
	\end{adjustbox}
	\caption{\small \textit{Multi-linear regression models with corresponding standardized $\beta$ coefficients. Dependent variables for each model are: episodic memory, verbal episodic memory, fluid intelligence ($gF$) and general intelligence ($g$).}}
\end{table}

\begin{table}[h]
	\centering
	\begin{adjustbox}{max width=\columnwidth}
		\begin{tabular}{|c|c|c|c|c|c|c|}
			\hline
			\multirow{2}{8.5em}{\bfseries MLM terms \\ /p-values}  & Constant & $\mathcal{P}^{FP}$ & $\mathcal{P}^{DMN}$ & $\mathcal{P}^{DA}$ & $\mathcal{P}^{SUBC}$& Entire \\\cline{2-6}
			& \bfseries	$p_0$ & \bfseries	$p_1$& \bfseries	$p_2$& \bfseries	$p_3$& \bfseries	$p_4$ & {\bfseries	Model} \\
			\hline\hline
			\bfseries Episodic Memory & 0 & 0.57 & 0.01 & & & 0.03\\
			\hline
			\bfseries Verb. Episodic Memory  & 0 & 0.02 & 0.77 & 0.17 & 0.03 & 0.04\\			
			\hline
			\bfseries $gF$ & 0 & 0.3 & $9\times 10^{-4}$ & & & 0.004 \\
			\hline
			$g$ & 0.03 & 0.44 & 0.16 & 0.57 & 0.05& 0.05  \\ 
			\hline\hline
		\end{tabular}
	\end{adjustbox}
	\caption{\small \textit{Multi-linear models with corresponding p-values. Note that we do not use step-wise linear model which discards descriptors that are not statistically significant. Column entire model shows the significance of the entire model.}}
\end{table}

The top panels in \textbf{Fig. 8} show  that as more linear descriptors (FN’s functional pre- and re-configurations) are added to iterative MLMs, variance associating with behavioral/cognitive performance measures decreases with linear descriptors that bear less subject sensitivity. This result highlights the importance of appending linear predictors in descending order with respect to the subject sensitivity. Specifically, as individual specificity reduces from left to right (\textbf{Fig. 7C}), the differential correlations, i.e. the difference between two consecutive correlation values, decreases.

\section{Discussion}
In this work, we fill a gap in the network neuroscience by proposing a mathematical framework that captures the extent to which subject-level functional networks, as estimated by fMRI, reconfigure across diverse mental/emotional states. We first propose that brain networks can undergo three different types of (re-)configurations: \textbf{i)} Network Configural Breadth, \textbf{ii)} Task-to-Task transitional reconfiguration, and \textbf{iii)} Within-Task reconfiguration. Unlike other existing frameworks \cite{schultz2016higher,shine2017principles,shine2018low,shine2019human}, the framework presented here can be applied to all three reconfiguration types.As a first step, in this study we focus on assessing the broadest aspect of reconfiguration,  i.e. Network Configural Breadth. We postulate, based on previous literature \cite{cole2014intrinsic}, that macro-scale (whole-brain) and micro-scale (edge-level) reconfigurations of brain networks are subtle, and hence difficult to disentangle. At the same time, mesoscopic structures in the brain (e.g. functional networks (FNs)) reconfigure substantially across different mental/emotional states as elicited by different tasks \cite{mohr2016integration}. The framework presented here constitutes the first attempt to formalize such (re)configurations of mesoscopic structures of the brain, and to quantify the behavior of a reference set of FNs with changing mental states. We set forth a novel, mathematically well-defined and well-behaved 2D network morphospace using novel mesoscopic metrics of Trapping Efficiency (\textbf{TE}) and Exit Entropy (\textbf{EE}). This morphospace not only characterizes the topology of FNs but also the flow of information within and between FNs. We show that this morphospace is sensitive to FNs, tasks, subjects, and the levels of cognitive performance.  We show that both of these measures are highly subject-sensitive for some FNs, while preconfiguration is highly subject-sensitive for all of them. Lastly, we also formalize and quantify the concepts of functional reconfiguration (the extent to which an FN has the capacity to reconfigure across different tasks) and functional preconfiguration (amount of transition from resting-state to a task-positive centroid). We thus construct a formalism that can explore FN changes across different cognitive states in a comprehensive manner and at different levels of granularity.
%10/11 6pm checkpoint

Ideally, a morphospace framework \cite{avena2015network,avena2018communication,corominas2013origins,goni2013exploring,mcghee1999theoretical,morgan2018low,schuetz2012multidimensional,shoval2012evolutionary,thomas2000evolutionary}
 would have a minimal complexity and, in this particular case, capture distinct features of functional network changes. As discussed in \cite{avena2015network}, metrics parametrizing a given morphospace should be disjoint. We see that, for any specific FN, high within-subject task sensitivity of \textbf{TE} does not necessarily imply a high value in \textbf{EE} and vice versa (e.g. VA and FP in \textbf{Fig. 4A}). In addition, we see that both \textbf{TE} and \textbf{EE} offer their unique insights in capturing non-overlapping features with \textbf{TE} is more subject-sensitive and \textbf{EE} more task-sensitive at the cohort level (\textbf{Fig. 4B,4C}). \textbf{Fig. 6B} highlights the disjoint nature of the two metrics as well, where we compute maximal distance per FN polytope in the TE and the EE axes separately. Results show that corresponding \textbf{TE} and \textbf{EE} maximal distances are disjoint and FN dependent. In other words, for a specific FN, the polytope is “stretched” in a particular task direction, where each morphospace measurement (\textbf{TE} or \textbf{EE}) unravels distinct properties. In \textbf{Fig. 6C}, we further see that a subset of tasks dominantly contribute to the maximal distance computation, such as Motion, Language, and Social tasks. Interestingly, we see that Motion and Language tasks can be considered “orthogonal” tasks with respect to \textbf{TE} and \textbf{EE}. %The fact that maximal distance offers different insights than hull volume (i.e. functional reconfiguration (\textbf{Fig. 7.1A})), also supports the idea that the minimal morphospace dimension is efficiently two.

Interestingly, the limbic network possesses the lowest ability to distinguish between tasks (\textbf{Fig. 4}). Similar behavior has been observed in Amico et al. \cite{amico2019centralized} when using Jensen-Shannon divergence as a distance metric of functional connectivity. In addition, the limbic network seems to work as a "relay" in brain communication \cite{amico2019towards}. One potential explanation for this unique behavior is that the limbic network maintains a minimal cognitive load across various tasks, most of which comprises relaying information from one part of the brain to the others; it thus does not reconfigure as much across different mental states.

Brain network configuration is typically studied considering a specific task at multiple spatial and temporal scales, see \cite{bassett2011dynamic,betzel2017positive,shine2016dynamics,shine2017principles,shine2018dynamic,shine2018low,shine2019human,mohr2016integration}. Previous investigations have mainly focused on the mechanism of how the brain traverses between high/low cognitive demands \cite{amico2019centralized,avena2018communication,bertolero2015modular,shine2018low,shine2019human,sporns2013network}, or on periods of integration and segregation at rest \cite{shine2017principles,shine2018dynamic,shine2018low,shine2019human}, defined in this paper as within-task reconfigurations. On the other hand, whole-brain configurations have also been investigated across different tasks (one configuration per task) with respect to rest, which led to the concept of general efficiency \cite{schultz2016higher}. This approach would belong to a wider category that we formally generalize as the Network Configural Breadth. The idea of general efficiency in \cite{schultz2016higher} relied on whole-brain FC correlations between task(s) and rest. While intuitive in quantifying similarity/distance between a single task and rest, quantification across multiple tasks becomes a challenge. Specifically, note that in Schultz and Cole \cite{schultz2016higher}, general efficiency is quantified using the first eigenmode, which explains most of the variance, after measuring the correlation between resting FC and three distinct task FCs. As more and more tasks are included, using the first eigenmode would become less and less representative of the task-related variations present in the data (in this paper summarized as the Network Configural Breadth). The proposed network morphospace overcomes these limitations and can be used to study brain network (re-)configurations across any number of tasks. It allows us to study different types of brain network (re-)configurations, as mentioned in above, using one comprehensive mathematical framework, which also facilitates a meaningful comparison between these seemingly disparate kinds of (re-)configurations. Schultz and Cole \cite{schultz2016higher} proposed that configurations can be compartmentalized into two differentiated concepts: functional reconfiguration and preconfiguration. Note that although the term \textbf{reconfiguration} is also used in \cite{schultz2016higher}, it is not referring to the action of switching among multiple mental/emotional states, i.e. as represented by task-to-task transitional reconfiguration or within-task reconfiguration (as shown in \textbf{Figure 1(b) and 1(c)}). Rather, it refers to the overall competence in exploring the total repertoire of task space of each subject given its resting configuration. That is why when we translate the corresponding idea into the mesoscopic morphospace, we call it the network configural breadth. We have also incorporated the two concepts of functional pre- and re-configurations into a well-defined mathematical space, which solves some of the technical difficulties (as discussed in \textbf{Section 2}) and generalizes these concepts to mesoscopic structures.

Brain network within-task reconfigurations have been almost exclusively qualitatively assessed. For instance, Shine et al. \cite{shine2016dynamics} show that the whole-brain functional connectome traverses segregated and integrated states as it reconfigures while performing a task. They also found that integrated states are associated with faster, more effective performance. Our formalism of within-task reconfigurations permits assessing such reconfigurations in a quantitative manner. Potentially, such within-task reconfigurations could also be used to assess cognitive fatigue, effort or learning across time.

Cole et al. \cite{cole2014intrinsic} have shown that the resting architecture network modifies itself to fit task requirements through subtle changes in functional edges. Numerically, small changes constituted by functional edges between rest and task-based connectivity might not be statistically significant when looking at edge level. Moreover, we also observe that while such changes might be negligible on a whole-brain global scale, they are more evident when looking at subsystems or functional brain networks, as clearly observed in the VIS network, relative to others. For functional preconfiguration (\textbf{Fig. 5, Fig. 6, Fig. 7.2A}), this effect is observable in all the FNs. In essence, we are postulating that a mesoscopic explorations of changes in brain network configurations with changing mental states is more informative than a macroscopic or microscopic exploration. 

%The proposed network morphospace does not study each mesoscopic structure in isolation: the mesoscopic structure (in our case FN) is not removed from the overall network for exploration. 

A key feature of this morphospace is that to study brain network (re-)configuration, an FN is not removed from the overal network for exploration. On the contrary, both metrics that define the morphospace, \textbf{TE} and \textbf{EE}, account for a particular FN's place embedded within the overall network, both in terms of topological structure and flow of information. That is why it is important to begin with a reference set of FNs (e.g., RSNs), so as to study how these FNs adapt to changing mental states within the context of the overall network.

Another benefit of a mesoscopic framework is that we can compare individual cognitive traits in each FN, instead of the whole brain (\textbf{Fig. 7.1B, 7.2B}). Specifically, after quantifying reconfiguration and preconfiguration for all FNs, we determine if these quantities incorporate information about individual traits (\textbf{Fig 7.C}). We observe different levels of subject fingerprint in different FNs for both re- and pre-configuration measures. This subject fingerprint heterogeneity across different FNs is consistent with previous literature on functional connectome fingerprinting, \cite{amico2018quest,finn2015functional}. Interestingly, functional pre-configuration (amount of transition from a resting-state to a task-positive state) displayed greater subject fingerprint than functional reconfiguration for all FNs. Based on this observation, we argue that to have better subject differentiability, we need to design tasks where the subject transitions from a stable resting-state to a task-positive state and/or vice versa \cite{amico2020disengaging}. This could be a significant step forward in precision psychiatry \cite{fraguas2016progressive}, where we can identify regional brain dysfunction more precisely as a function of the type and degree of cognitive or emotional load.

Subject-sensitivity of the proposed network morphospace framework is also supported by significant associations of the frontoparietal and default mode networks with fluid intelligence, see \textbf{Table 1} \& \textbf{Table 2}. Specifically, as pointed out by Tschentscher et al. \cite{tschentscher2017fluid}, high fluid intelligence is associated with a greater frontoparietal network activation, which is also consistent with findings from a three-back working memory task (Gray et al. \cite{gray2003neural}). In the domain of network configural breadth, we observe a higher reconfiguration as represented by a positive frontoparietal functional preconfiguration coefficient (\textbf{Table 1}). 

%The level of FP network involvement is an indicator of its corresponding distance from rest to the task-positive position (geometric centroid) in the morphospace.

%Another significant contribution of this work is that it confirms the associations of frontoparietal and default mode networks with fluid intelligence. Specifically, as pointed out in Tschentscher et al. \cite{tschentscher2017fluid}, high fluid intelligence corresponds with the greater frontoparietal network activation, consistent with findings from the three-back working memory task (Gray et al. \cite{gray2003neural}). In the domain of network configural breadth, this is demonstrated by a positive frontoparietal preconfiguration coefficient. The level of FP network involvement is an indicator of its corresponding distance from rest to the task-general position (geometric centroid) in the morphospace. On the other hand, default mode network efficiency (based on normalized path length) correlates negatively with IQ (van den Heuvel et al. \cite{van2009efficiency}). Analogously, when viewed under network configural breadth, DMN preconfiguration is also inversely associated with g and gF (see Table 1). Comprehensively, an interesting observation drawn from the selected multi-linear model is that, on the global scale, brain reconfiguration also correlates negatively with intelligence (Schultz and Cole \cite{schultz2016higher}). However, when investigating the mesoscopic structures, i.e. functional networks, this observation cannot be generalized to all functional communities.

This study has several limitations. The framework was tested specifically on the Human Connectome Project dataset and using a single whole-brain parcellation. Alternative parcellations \cite{schaefer2018local,tian2020hierarchical}, additional fMRI tasks to better sample the cognitive space, and other datasets might offer further insights about the mesoscopic network morphospace, see \cite{corominas2013origins,avena2015network}. We did not perform a sensitivity analysis on how small fluctuations in functional connectomes affect mapping into the network morphospace. Due to the nature of module trapping efficiency and exit entropy metrics, negative functional couplings were not considered and hence, were set to zero.

% multimodal parcellation proposed by Glasser et al. [30] and subsequently organized the regions into seven cortical FNs (as per Yeo et al. \cite{yeo2011organization}) and a sub-cortical network. Alternative parcellations of the human brain, task selection, and subject populations might offer different insights. A second limitation is the sensitivity of computed points in mesoscopic morphospace per connectome construction. Another consideration is that negative functional couplings are not considered due to the nature of module trapping efficiency and mode exit entropy measurements. In the context of trapping efficiency measurement, having negative functional edges does not support subsequent construction, especially after it is ranged in the non-negative domain. The fourth consideration is that the quantified configural breadth terms, although not rigorously true to scale, are fairly well-calibrated. The evidence for this is Fig. 6B when we compute the maximal distance in VIS and DA networks in which both measures have similar magnitude. Hence, if visual network reconfiguration is twice (in numerical sense) larger than limbic reconfiguration, although we cannot simply infer strict numerical relationship based solely on this result, we can reasonably said that VIS reconfigures “more” than LIM. 

Future studies should incorporate a sensitivity study of the behavior of this network morphospace with respect to small fluctuations in the input functional connectomes. Further studies could also incorporate structural connectivity information to inform both \textbf{TE} and \textbf{EE} measures when assessing the morphospace coordinates of functional reconfiguration. Additional exploration of different aspects of this morphospace could provide further insights. For example, location of the polytopes in the morphospace might improve individual fingerprint. An important aspect of the proposed mesoscopic network morphospace is that it allows for an exhaustive and continuous exploration of network reconfigurations, including those that are continuous in time \cite{douw2016state,shine2018low}. For example, if the subject performs several tasks within the same scanning session, including extended resting-state periods (such as the fMRI experiment done at \cite{barnes2009endogenous}. This would allow us to fully explore the cognitive space and gain a valuable insight into how different subjects adapt to different levels of cognitive demands. One can also study the trajectory of changing mental states using dynamic functional connectivity (Gonzalez-Castillo et al. \cite{gonzalez2015tracking}), which can easily be mapped to this morphospace for additional insights. Another potential avenue could be the application of this framework to characterize and understand different brain disorders.

%Future studies should incorporate a sensitivity study of the behavior of those points $u(\mathcal{ C})$ with respect to the input connectomes. Specifically, future studies can focus on this space behavior with the introduction of, possibly, Gaussian noise into functional connectomes and re-analyze the results. One should also check the location of those polytopes in this space to see if there exists any individual fingerprints in such space. Another important aspect of the mesoscopic morphospace is that it allows for exhaustive and continuous exploration of network shifts, including those that are continuous transitions in time, \cite{douw2016state,shine2018low}. Hence, future studies could also involve the first two types of reconfigurations: within-task and task-to-task \cite{barnes2009endogenous}. For a specific task, one can study network shifts between integrated and segregated brain states using the mesoscopic morphospace. Hence, one can also study the trajectory of resting periods using dynamic functional connectivity (Gonzalez-Castillo et al. \cite{gonzalez2015tracking}). In this quest, the morphospace is a well-suited tool to accomplish such aim. Finally, another possible extension could incorporate structural connectivity information so that only functional edges supported by the presence of white-matter streamlines would be assessed.

In summary, this mesoscopic network morphospace is our first attempt to create a mathematically well-defined framework to explore an individual's cognitive space at different levels of granularity. It allows us to characterize the structure and dynamics of specific subsystems in the brain. We show that this morphospace is sensitive to specific FNs, cognitive states, individuals, and levels of cognitive performance in different tasks. We formally define three different types of (re-)configurations that an individual's brain can traverse and provide a method to quantify the resulting changes in brain network organization. This type of framework can be extremely helpful in characterizing brain dynamics at individual-level, in healthy and pathological population, which in turn would pave the way for the development of personalized medicine for brain disorders.

%In summary, this work has established a unified method for quantifying the exploratory capacity within the subject’s cognitive space considering various mental states, and its associations with measures of intelligence. Future work can leverage this approach to studying how networks are affected by individual characteristics and disease.

%\newpage
\begin{singlespace}

\section*{Methodology}
We provide the detailed information on materials and methods in \textbf{SI}. In short, all necessary mechanics collected from multiple disciplines and general set-up for matrix computations are described \textbf{SI} Preliminaries and Methods. The data set is consisted of high-resolution functional connectivity matrices describing human cerebral cortex and sub-cortex (see \textbf{SI} Data Description). The construction of morphospace and the formalized notion of configural breadth are described in \textbf{SI} Morphospace analysis and configural breadth section. Multi-linear model and model specificity are described in \textbf{SI} Behavioral Measure Analysis. 

\section*{Acknowledgments}
Data were provided [in part] by the Human Connectome Project, WU-Minn
Consortium (Principal Investigators: David Van Essen and Kamil Ugurbil; 1U54MH091657) funded by the 16 NIH Institutes and Centers that support the NIH Blueprint for Neuroscience Research; and by the McDonnell Center for Systems Neuroscience at Washington University. JG acknowledges financial support from NIH R01EB022574 and NIH R01MH108467 and the Indiana Clinical and Translational Sciences Institute (Grant Number UL1TR001108) from the National Institutes of Health, National Center for Advancing Translational Sciences, Clinical and Translational Sciences Award. MV and JG acknowledges financial support from Purdue Industrial Engineering Frontier Teams Network Morphospace Award and from Purdue Discovery Park Data Science Award "Fingerprints of the Human Brain: A Data Science Perspective." We thank Dr. Olaf Sporns and Meenusree Rajapandian for valuable comments.

\section*{Author Contributions} 
D-T.D., J.G., E.A. designed research; D-T.D. and J.G. performed research; D-T.D., J.G., K.A., E.A., C.-M.B., M.V., M.D., D.K. contributed new reagents/analytic tools; D.-T.D., J.G., K.A., E.A. analyzed data; and D.-T.D and J.G. wrote the paper with K.A, E.A., C.-M.B., M.D., D.K. and M.V. providing comments and edits. 

\section*{Author Declaration}
The authors declare no conflict of interest.
\end{singlespace}

%\newpage
\section*{Supplemental Information (SI)}
This document purpose is to collaborate on the machinery of the morphospace and other aspects such as the data set and brain atlas used to analyze the data. The aim is to provide further analytic results in conjunction with the ones that are already presented in the main paper. 
\appendix
%\counterwithin{figure}{section}
\renewcommand{\thefigure}{S\arabic{figure}}
\setcounter{figure}{0}
\section{Preliminaries}
In this section, we establish some of the key mathematical notations used throughout the paper. Specifically, scalar is italicized, $a$. A vector is denoted as bold letter, $\mathbf{a}$ (Default mode is in column fashion). Matrix is notated as capitalized bold letter, $\mathbf{A}$.  If $r\in [q]$ where $q\in \mathbb{N}^{+}$, it means that $r$ accepts integer values from 1 up to, including, $q$. Given any set $S$, its cardinality is denoted as $|S|$. Given any two vectors $\mathbf{a,b}\in \mathbb{R}^{n}$, $\langle \mathbf{a},\mathbf{b}\rangle$ denote inner product.
In terms of graph theory, a weighted network is denoted as $G(V,E)$ where $V$ and $E$ are sets of vertices and edges in such network, respectively. $G(V,E)$ can be represented by $\mathbf{A}_{G}=\mathbf{A}(ij)=[w_{ij}]$, in which $w_{ij} \in [0,1]$ represents functional coupling strength between node $i$ and $j$. The strength of node $i\in V(G)$ is denoted as $k_i$, typically stored in the diagonal matrix $\mathbf{K}$ for $\mathbf{K}(ii)=k_i$. A generic matrix $\mathbf{F}$ with entries valued in a continuous interval $[x,y]$, $z_1$ rows and $z_2$ columns is denoted as $\mathbf{F}\in [x,y]^{z_1\times z_2}$. Further, if we want to induce a sub-matrix from the original matrix $\mathbf{F}$ based on a specific set of rows, denoted as set $S_{rows}$, and columns, denoted as set $S_{columns}$, we use the notation: $\restr{\mathbf{F}}{S_{rows},S_{columns}}$. If the set of rows and columns are matched (both denoted as $S_{w}$), then we will ease notation by using $\restr{\mathbf{F}}{S_w}$.

\section{Data}
In this section, we provide the details related to the dataset we used to analyze the notion of configural breadth. We also provide information related to the brain atlas.  

\subsection{Brain atlas}
The brain atlas used in this work is the based on the cortical parcellation of 360 brain regions as recently proposed by Glasser et al. \cite{glasser2016multi}. Similarly to reference \cite{amico2018quest,amico2018mapping}, 14 sub-cortical regions were added, as provided by the HCP release (filename $Atlas$\_$ROI2.nii.gz$). We accomplish this by converting this file from NIFTI to CIFTI format by using the HCP workbench software [\url{(http://www.humanconnectome.org/software/connectome workbench.html}, with the command -cifti- create-label]. This resulted in a brain atlas of 374 brain regions (360 cortical + 14 sub-cortical nodes). 

Using Human Connectome Project Dataset, we explore the characteristics of functional networks' configural breadth by utilizing Resting State Networks (FNs), see \cite{yeo2011organization}, which includes seven functional networks (FNs): Visual (VIS), SomatoMotor (SM), Dorsal Attention (DA), Ventral Attention (VA), Limbic (LIM), FrontalParieto (FP), Default Mode Network (DMN); Sub-cortical (SUBC) region, as mentioned before, is added into this atlas for completeness. Thus, the parcellation used in this paper comprises of eight (8) FNs.

\subsection{HCP Dataset}
The fMRI dataset used in this paper is available in the Human Connectome Project (HCP) depository (\url{http://www.humanconnectome.org/}), with Released Q3.  The processed functional connectomes obtained from this data and used for the current study are available from the corresponding author on reasonable request. Please refer to below detailed descriptions on the dataset and data processing.

\subsection{HCP Functional Data}
The fMRI data from the 100 unrelated subjects in the HCP Q3 release were employed in this study \cite{van2012human}, \cite{van2013wu}. Per HCP protocol, all subjects gave written informed consent to the HCP consortium. The two resting-state functional MRI acquisitions (HCP filenames: $rfMRI\_REST_1$ and $rfMRI\_REST_2$) were acquired in separate sessions on two different days, with two distinct scanning patterns (left to right and right to left) in each day, \cite{glasser2013minimal}, \cite{van2012human}, and \cite{van2013wu} for details. This release includes also data from seven different fMRI tasks: gambling ($tfMRI\_GAMBLING$), relational or reasoning ($tfMRI\_RELATIONAL$), social ($tfMRI\_SOCIAL$), working memory ($tfMRI\_WM$), motor ($tfMRI\_MOTOR$), language ($tfMRI\_LANGUAGE$, including both a story-listening and arithmetic task), and emotion ($tfMRI\_EMOTION$). Per \cite{glasser2013minimal}, \cite{barch2013function}, three tasks MRIs are obtained: working memory, motor, and gambling. 

The local Institutional Review Board at Washington University in St. Louis approve all the protocol used during the data acquisition process. Please refer to \cite{glasser2013minimal,barch2013function,smith2013resting} for further details on the HCP dataset. All tasks and resting functional MRIs are equally weighted importance. In other words, no particular weight is assigned to any specific tasks. 

\subsection{Constructing functional connectomes}
We used the standard HCP functional pre-processing pipeline, which includes artifact removal, motion correction and registration to standard space, as described in \cite{glasser2013minimal,smith2013resting} for this dataset. For the resting-state fMRI data, we also added the following steps: global gray matter signal regression; a bandpass first-order Butterworth filter in both directions; z-scores of voxel time courses with outlier eliminations beyond the three standard deviations from first moment \cite{marcus2011informatics,power2014methods}. 
\begin{figure}[H]
	\centering
	{\includegraphics[width=.9\columnwidth]{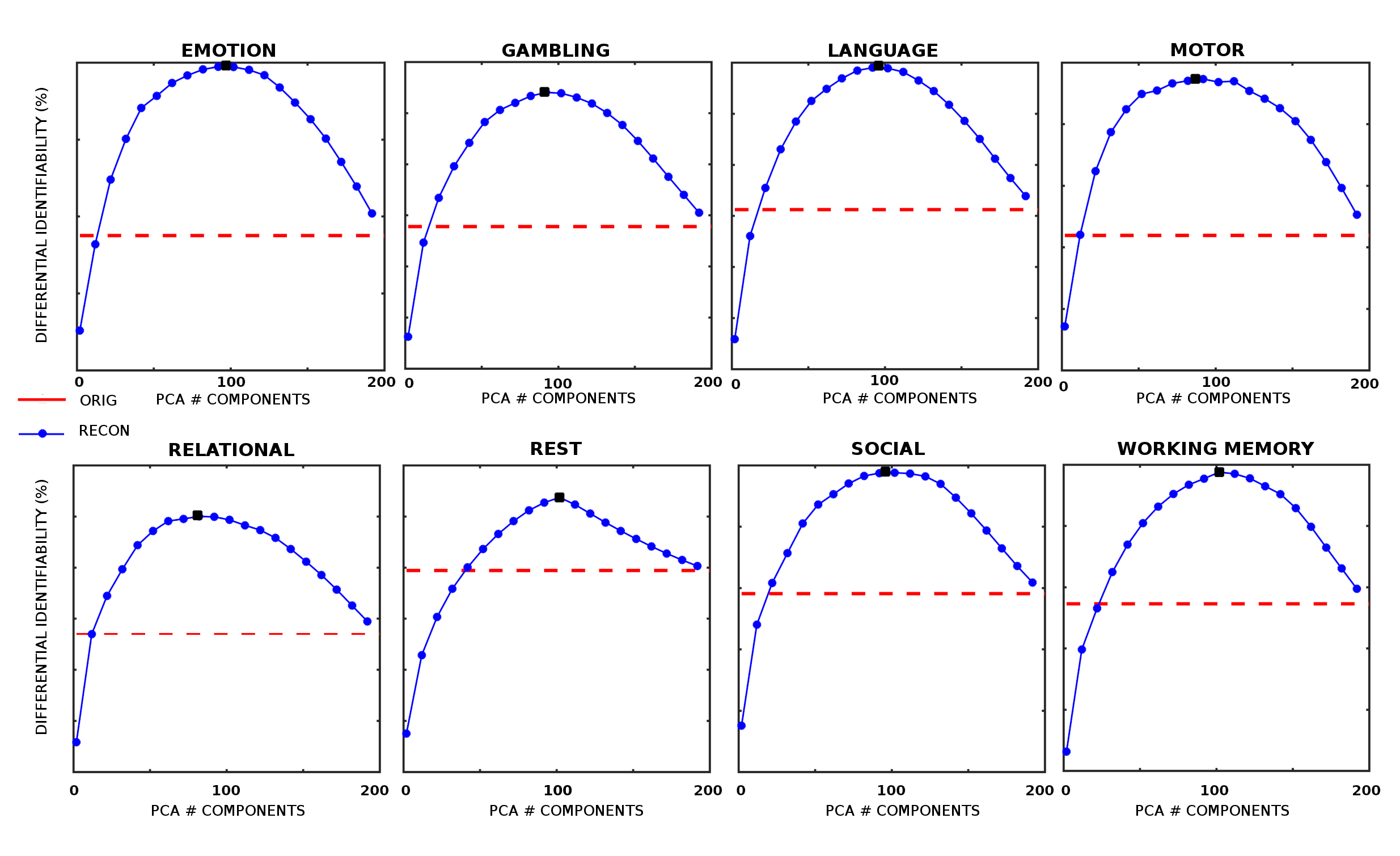}\label{fig:PCA_recon}}
	%	\newline
	\caption{\small The framework proposed by Amico et al. \cite{amico2018quest} is used to maximized individual fingerprints. Each subplot represent rest and seven tasks in HCP dataset. The optimal reconstructed number of orthogonal components is indicated by a black dot.}
\end{figure}
For task fMRI data, aforementioned steps are applied, with a relaxation for bandpass filter [0.001 Hz, 0.25 Hz]. Starting from each pairs of nodal time courses, Pearson correlation is used to fill out the functional connectomes for all subjects at rest and seven designated tasks. This would yield symmetrical connectivity matrix for all fMRI sections.

\textit{Resting State Connectomes:} There are two resting scanning sections conducted in two different days. In each day, individual MRIs are obtained independently in the morning and afternoon sections. We average the resting functional connectome in the first day (which contains morning/afternoon scans) and call it Test. By the same token, we obtain Retest FC for resting condition. 

\textit{FC's matrix entries:} For all considered fMRI images in this paper, we first threshold negative correlations. This is purely technical because one of the morphospace axis is built upon stochastic ground; hence, numerically it is not possible to utilize negative entries.   The remaining matrix are, then, squared. 

\textit{Improve Individual fingerprint}
To improve identifiability in human functional connectome, we utilize data dimensional reduction technique described in Amico et al. \cite{amico2018quest}, see \textbf{Figure S1}.

\section{Morphospace Analysis}
The concept of a morphospace can be used to analyze many other mathematical objects, including networks. When applied to networks, quantitative traits of global or local network topology are conceptualized through the Cartesian coordinates defined in this abstract space. A brain’s subsystem configuration is topologically represented by a point in this multidimensional space. Constructing a morphospace provide us the freedom formally define and track desirable phenotypes of the under-studied objects. In this section, we analyze both measurements in depth. We first introduce the formulation of each axis and then provide further characteristics and/or requirements/assumptions, if any. Any theory that already introduced in the main text will not be re-introduced here.   

It is important to be aware that any generic network can be fragmented (disconnected). This is rarely the case for brain functional networks because of how we compute functional couplings, typically using Pearson Correlation Coefficient. Despite of that, \textit{a priori} functional community induced from global thresholded adjacency structure is  not guaranteed to be connected. As pointed out in \cite{malliaros2013clustering} among others, a meaningful cluster should, at minimum, be connected. To respect the global topology of the functional networks $G$, i.e. connectedness, we applied a small perturbation to edges with zero weight, i.e. 
$$a_{ij}=0\rightarrow a_{ij}=\eps\mid\forall i,j\in \mathcal{C}$$
According to \cite{edelsbrunner2010computational}, if one concerns solely on the connectedness property, then topological spaces are similar to graphs. Therefore, the goal is to have the connectedness property carried from the graph $G$ to all of its induced subgraphs $\mathcal{C}$. To do so, one needs to preserve the topology defined on $G$, i.e. the collections of open sets $\mathcal{U}$ (which can be thought of as the edge set $E$ defined on $G$).

\subsection{The Coordinates of mesoscopic morphospace}
In this section, we describe the mesoscopic morphospace Cartesian axes in greater detail. The two coordinates are Module Trapping Efficiency ($\mathbf{TE}$) and Exit Entropy ($\mathbf{EE}$). Thus, for any functional module, $\mathcal{C}\subset G$, we define a point in morphospace $\Omega$ to be
\[
u(\mathcal{C}) = (\mathbf{TE}(\mathcal{C}), \mathbf{EE}(\mathcal{C}))
\]

\subsection{Module Trapping Efficiency}
This is the $x$-coordinate of $u(\mathcal{C})$.\\ Module trapping efficiency assesses the characteristic of a functional community based on how well it sustains its topology under rich repertoire of task-evoked conditions, relatively to its segregation/integration role, simultaneously. Recall that module Trapping efficiency is formalized as followed:
\[
\mathbf{TE}(\mathcal{C})=\frac{||\tau||_2}{\mathcal{L}_\mathcal{C}}
\]
\subsubsection{Numerator $\tau$}
As claimed in the main text, $\mathbf{TE}$ is finitely bounded. There are several ways to observe this; one approach involves applying hierarchical community detection algorithm \cite{fortunato2016community} and look for the first time $G$ split into more than one subgraphs. Thus, let $i$ be indices representing communities belong to the first hierarchical layer, then $$M=\max_{k} [\mathbf{TE}(S_k)]\mid \forall k \in [l]$$ where $l$ represents the number of communities.
\begin{figure*}
	\centering
	\label{Fig:TE_upperbound}
	\subfloat[Edge strength, after post-processing steps, histogram of all considered FCs. ]{\includegraphics[width=.47\columnwidth]{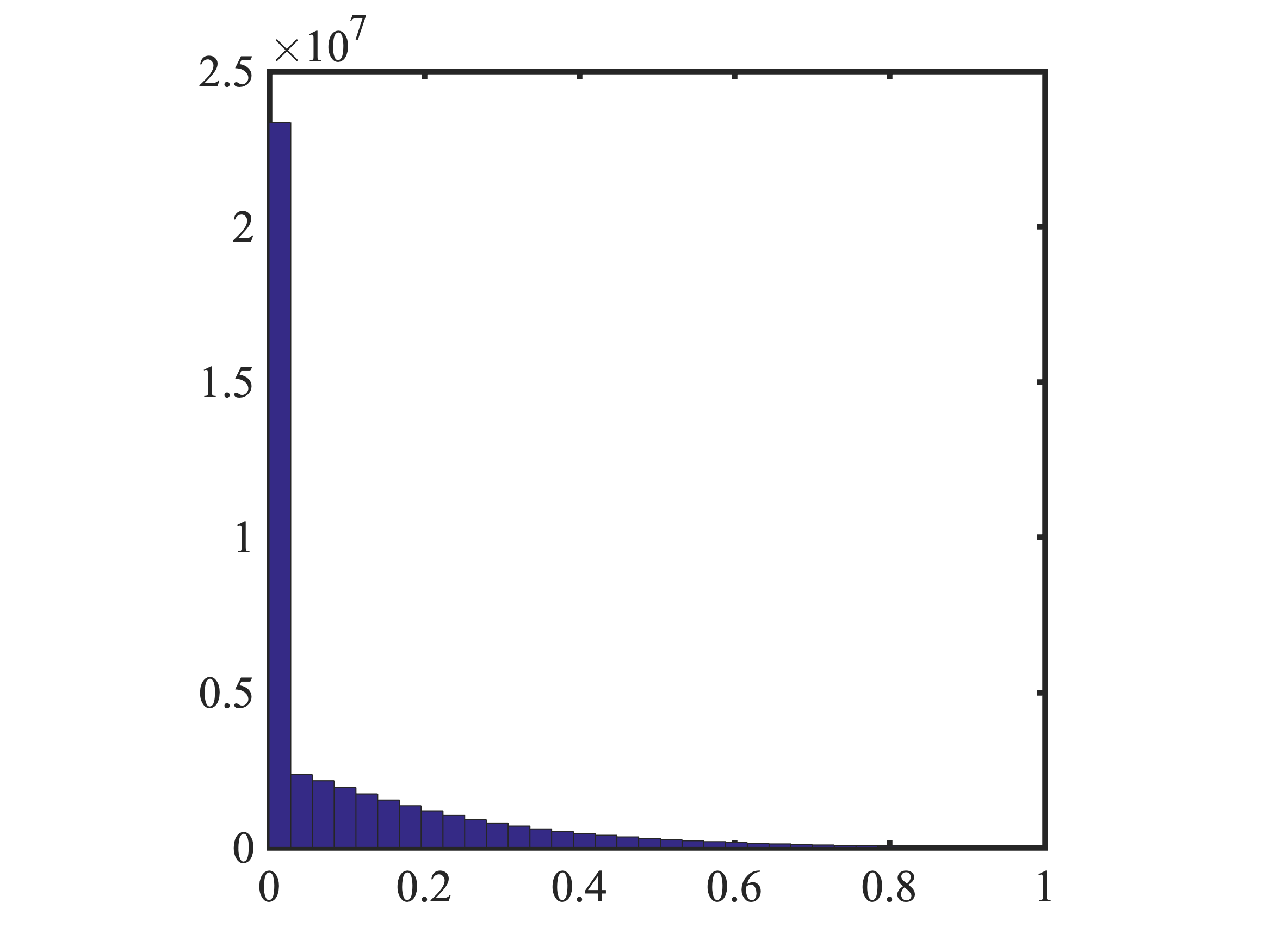}}
	\hspace{2mm}
	\subfloat[FC density, defined to be the number of non-zero functional weighted edges out of $N\choose 2$ possible edges, histogram of all considered FCs. ]{\includegraphics[width=.47\columnwidth]{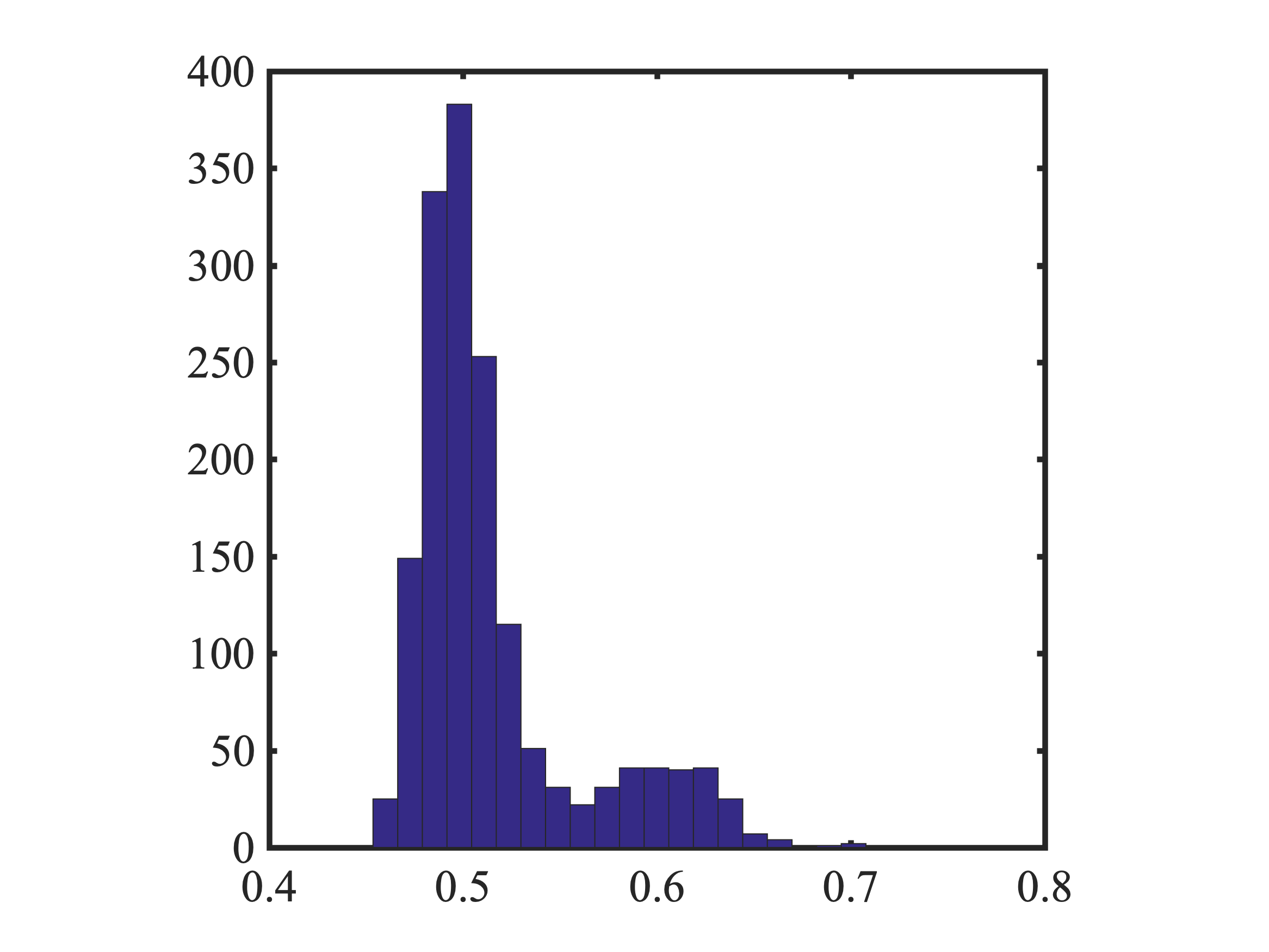}}
	\caption{Mean Density and majority of edge strength falls in the first bin [0,0.025] are the two major factors into the maximum value of \textbf{TE}.}
\end{figure*}
Such value is well-defined and finite. An alternative way to see the trivial bound of the measures is as follows: Let us consider the entire network $G$, we have:
\[
\mathbf{TE}(\mathcal{ C}\equiv G) = \frac{||\tau||_2}{\mathcal{L}_\mathcal{C}} = \frac{\infty}{0} = \infty
\]
\footnote{Note that $\frac{\infty}{0}$ is undefined. However, in such case, we define this quantity to be unbounded which is the notion of infinity.}because there is no exits if the configurations is the entire network; moreover, there is zero leakages. Hence, any cut into $G$ would have to be strictly less than this upper bound. 

In the context of the data set at hand, we can, however provide a better bound then finiteness. We proceed by obtaining the maximum value of \textbf{TE} when all subjects and all tasks are under consideration which yields the result 
\[\max_{subjects,tasks}(\mathbf{TE}) = 0.5064\]
One can relate this numerical value with two factors: Functional connectome density and edge strengths, see \textbf{Fig. S2} for further details.

\subsubsection{Normalization}
Realistically, since larger communities carry more exits which is driven purely from a topological viewpoint, $\mathcal{L}_\mathcal{C}$ is a logical choice to normalize the magnitude of $\tau$. 

Additionally, $\mathcal{L}_\mathcal{C}$ is deemed to perform as $||\tau||_2$-damping. Notice that there also exists functional communities with low total exiting strength with large cardinality, theoretically. In such case, these structures are rewarded from the standpoint of $\mathbf{TE}$ as it converges to $\mathbf{TE}(\mathcal{C}\equiv G)$.

\subsection{Module Exit Entropy}
This is the $y$-coordinate of $u(\mathcal{C})$.\\ Module exit entropy represents communicating preferences of $\mathcal{C}$ with respect to the rest of network $G$ from information theoretical viewpoint. The magnitude of this measure assesses the specificity of integration of a given community in varied environment. 
\[
\mathbf{EE} =\frac{\mathcal{H}_e}{\mathcal{N}_{\mathcal{C}}} = -\frac{\langle \psi, log(\psi)\rangle}{log(|S_{abs}|)} = \frac{-\sum_{i=1}^{|S_{abs}|} \psi_i \log(\psi_i)}{\log (|S_{abs}|)}
\]
where $\psi^T = [\psi_i]_{i=1}^{i=|S_{abs}|} =\mathbf{1}^T_{|V_\mathcal{C}|} \Psi\sbrac{\mathbf{1}^T_{|V_\mathcal{C}|} \Psi \mathbf{1}_{|S_{abs}|}}^{-1} $.

\subsubsection{Numerator}
The numerator of $\mathbf{EE}(\mathcal{C})$, i.e. $-\sum_{i=1}^{|S_{trans}|} \psi_i log(\psi_i)$, measures the extent to which specified channels of communications, under finest scale (i.e. node/edge-level), is established between nodes in $\mathcal{C}$ with nodes that belong to other functional communities in $G$. Therefore, $-\sum_{i=1}^{|S_{trans}|} \psi_i log(\psi_i)$ does not necessary dependent on the connectivity strength, represented by $w_{ij}\forall i\in\mathcal{C},j\in\mathcal{J}$. For example, let us say that we have two communities with the same state set $S_{trans}={1,2,3}$ and $S_{abs}={a,b,c,d}$. In community 1, $w_{ij}=0.01,\forall i\in S_{trans},j\in S_{abs}$; and community 2,  $w_{ij}=0.9,\forall i\in S_{trans},j\in S_{abs}$. Once we compute the normalized entropy, i.e. \textbf{EE}, for both cases, we see that they both have no communication preference, hence $\mathbf{EE}=1$ for both cases.
%\subsubsection{Numerical Range}
%The maximum value it can take on is one, which represents no particular preference from $\mathcal{C}$ to $G\setminus \mathcal{C}$. Note that, conceptually, this is not the same as $\mathcal{L}_S$ incorporated in $\mathbf{TE}(\mathcal{C})$ (as discussed later) as $\mathbf{EE}(\mathcal{C})$ is driven solely from information theoretical viewpoint of the module with respect to its external topology. On the other hand, numerically, the value towards 0 demonstrates extreme preference in terms of integrative duties. It means that nodes in $\mathcal{C}$ have very specific nodes outside of $\mathcal{C}$ such that established channels of communication take place.

\subsubsection{Normalization}
This is the coordinate where normalization is possible. Note that since entropy is normalized by its maximum value (i.e. $log(|S_{trans}|)$), the number of exits $|S_{trans}|$ impacts is, consequently, neutralized. Thus, one does not need to concern about the cardinality of a community with respect to its number of exits as, in reality, a typically larger community usually carries more exits.

\subsection{Morphospace Null Model}
\subsubsection{Randomization of $G(V,E)$}
Given a weighted network $A= [w_{ij}]$, we apply randomized algorithm \url{Xswap}, , see \cite{hanhijarvi2009randomization} with number of desired changes that are set to be $[2, 2^3, 2^5 ... ,2^{19}]$ (with exponent increment of 2) and maximum iterations set at 100 times the corresponding changes. This algorithm \textbf{preserves network basic topological characteristics} such as size, density and degree sequence. As the desired number of changes increases, the difference between the original matrix, denoted as $\mathbf{A}_{orig}$, and the randomized counterpart, denoted as $\mathbf{A}_{rand}$, also increases. The difference between two graphs can be quantified as follows:
\[
Diss= \frac{\sum_{i,j=1}^{n}\left| \mathbf{A}_{rand} (ij) - \mathbf{A}_{orig} (ij) \right|}{\sum_{i,j=1}^{n} \mathbf{A}_{rand}(ij)}
\]
where $n$ is graph's size and $Diss \in [0,1]$. It is important to note that the difference between two graphs saturates after a certain number of changes and each graph topology saturates at different values (not necessarily $1$). 

Hence, getting $Diss$ to arbitrarily close to saturation with the smallest number of changes is genuinely the target for this procedure, see \textbf{Fig. S3}-Panel \textbf{A} for details. In our case, we pick subject 100307 and run the randomization procedure for all available tasks and rest. We first found that the acceptable $Diss$ occurs at $2^{15}$ desired changes at resting state. We note that the $Diss$ saturates at 0.6 because the (sub)graphs we are dealing with are very dense and some links will be repeated in force, leading to a non-zero overlap between the links of the graphs in the random ensemble and the original one. We then used the same number of changes for the investigated tasks in this dataset.

%\begin{figure}[h]
%	\centering
%	{\includegraphics[width=.6\columnwidth]{FigureS6.png} \label{fig:subject1_dissimilarity_atrest}}
%	\caption{\small \textbf{Connectome Randomization:} Subject 100307 resting state FC are randomized with the aforementioned step number of changes and the corresponding dissimilarity indices between the original graph and the randomized one. Due to computational demand of the produce, we only run \url{Xswap} for resting FC and choose the acceptable number of changes to be $2^{15}$. We apply the same number of changes to four considered tasks to demonstrate the trajectory of points. In the above graph, $x-$ axis represent the logarithmic scale of effective xswaps, denoted as $x$, operations and $y-$ axis represents the dissimilarity between the original graph and the randomized counterpart.}
%\end{figure}

\subsubsection{Morphospace Null Model}
The main drive for studying trajectories, through randomization, is because it could provide further evidence of the robustness in design of the metrics. Specifically, if done correctly, measurements should highlight unique characteristics of functional communities (and not the randomized counterpart). As the randomized graph (with topological preserved features) get assigned the same partition into functional networks (e.g. Yeo's parcellation) as the original one, any destruction of such topology, at global scale $G$, would also be carried over (hence, identified) by the morphospace itself. The result is shown in \textbf{Fig. S3}. One common theme emerges is that regardless of which functional network and task, as the dissimilarity increases with the desired number of changes, all functional communities are pushed towards to top left corner. This regime of the morphospace represent random exiting strategy from module $\mathcal{C}$ (high value of $\mathbf{EE}$ - high level of uncertainty in communication preference) and high degree of non-assortative community (low $\mathbf{TE}$ - low level of segregation). 

\begin{figure}[H]
	\centering
	{\includegraphics[width=\columnwidth]{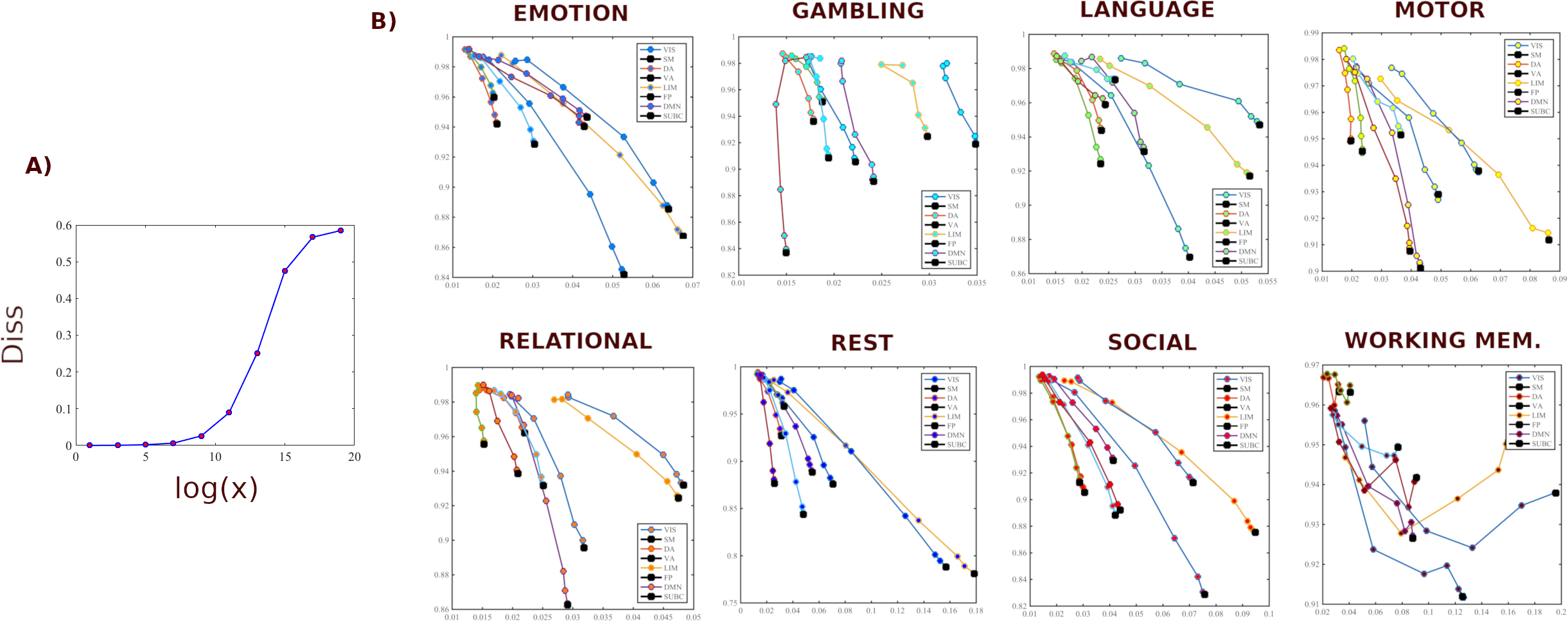}\label{fig:Morpho_trajectory}}
	\caption{\small \textbf{Morphospace Null Model}: Panel \textbf{A)} Subject 100307 with resting connectome using \url{Xswap} procedure. Panel \textbf{B)} represent the application of the selected number of changes to all other tasks. This is an important results to see that functional networks' topology is truly well-defined and highly reproducible across subject domain. Note that black square dot denote functional community \textbf{TE} and \textbf{EE} with no randomization. Color available online.}
\end{figure}

\section{Network Configural breadth}
\subsection{Polytope Theory}
Given a set of points $W=\cbrac{x_1,x_2,....x_{|W|}}$ for which $x_j\in\mathbb{R}^d,\forall j\in [|W|]$, a convex hull formed by such set of points are mathematically represented by 
\[
\mathbf{Conv}(W)= \cbrac{\sum_{j=1}^{|W|} \a_j x_j \mid \sum_{j=1}^{|W|} \a_j=1, \a_j\ge 0,\forall j\in [|W|]}
\]
where $d$ is called the ambient space dimension. Moreover, if $|W|\ge d+1$, we recall that points in $W$ are in general position if no hyperplane, i.e. flat of dimension $d-1$ contains more than $d$ points, \cite{ziegler2012lectures}. Otherwise, i.e. $|W|\le d$, there exist(s) point(s) that are affinely dependent on other points in $W$. 

%Given the convex hull induced by points $x_i$ in $W$, the convex hull dimension is defined to the be geometrical dimension of the polytope formed by points the hull. The convex hull dimension, denoted as $h$, is governed largely by the number of points participating, i.e. $|W|$. Given points $x_i\in W$, these points belong either to
%\begin{itemize}
%	\item the boundary (Pareto front) - sometimes, these points are referred to as vertices of the hull \cite{ziegler2012lectures};
%	\item the interior of the hull
%\end{itemize}
%
%For each pair of points that form the Pareto front (i.e. vertices of the hull), let us define two type of point pairs as follows: 
%\begin{itemize}
%	\item type A pairs are such that their convex combination belongs to the boundary of convex hull, denoted as $\delta (\mathbf{conv}(W))$;
%	\item type B pairs are such that their convex combination belongs to the interior of hull $W$, denoted as $int (\mathbf{conv}(W))$
%\end{itemize}

Providing that points in $W$ in $\mathbb{R}^d$, the approximated volume induced by the convex hull $\mathbf{Conv}(W)$ can be calculated through the formation of Delaunay Triangulation process \cite{ziegler2012lectures}.  
%Note that the volume of a convex hull depends on its dimension $h$ which is upper-bounded by the dimension of the ambient space $d$. 
The volume of the convex hull is denoted as $\mathbf{Vol}(\mathbf{Conv}(W))$. In $\mathbb{R}^d$, the convex hull dimension can take on the values
\begin{enumerate}
	\item $h=0$ which constitutes a point in $\mathbb{R}^d$, $\mathbf{Vol}(\mathbf{Conv}(W))=0$
	\item $h=1$ which constitutes a line segment, $\mathbf{Vol}(\mathbf{Conv}(W))= \sup (d(x_i,x_j)),\forall x_i,x_j\in W$ where $d(x_i,x_j)$ denotes the pre-defined metric distance between two generic points.
	\item $h=2$ and $h\ge 3$ which constitutes the notion of area and volume, respectively.
\end{enumerate}
\begin{figure*}[h]
	\centering
	\label{fig:convexdemo}
	\includegraphics[width=.8\columnwidth]{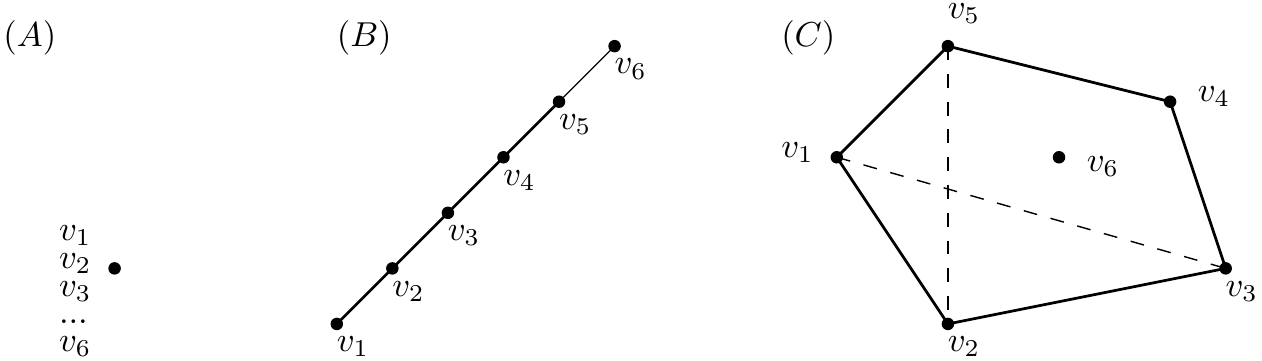}
	\caption{\small Given that $W=\cbrac{v_1,v_2,...,v_6}$, we demonstrate three possible scenarios of convex hull formed by $W$ in morphospace $\Omega$. Case \textit{(A),(B),(C)} correspond to the polytope dimension of $h=0,1,2$, respectively. Here we see that $\cbrac{v_1,v_6}$ and $\cbrac{v_1,v_2,v_3,v_4,v_5}$ forms the Pareto front in Case $(B)$ Case $(C)$, respectively. In case $(C)$, $v_6$ belongs to the interior of the hull. Further, in case (B) and (C), we see that the hull vertices, i.e. points belong to the Pareto front of the hull, are $\cbrac{v_1,v_5}$ for case (B) and $\cbrac{v_1,v_2,v_3,v_4,v_5}$ for case (C). Given the nature of this space, the first two scenarios are statistically rare. In the third scenario, we see that all 5 points constitute the boundary of $\mathbf{conv}(W)$. Further, we see that some type A pairs of points, graphically represented by solid lines, are $(v_1,v_5),(v_2,v_3)$ while some type B pairs, represented by dashed lines, are $(v_2,v_4),(v_3,v_5)$. }
	\label{fig:reconfig_demo}
\end{figure*}
For $h\ge 2$, convex hull volume is calculated using \url{Qhull} package implemented in Matlab, see \cite{Qhull}. In general, as pointed out also in \cite{Qhull}, computing $\mathcal{V}-$ or $\mathcal{H}-$ polytope metric volume is NP-hard (see also \cite{dyer1988complexity}, \cite{khachiyan1993chapter}) with the availability of efficient approximating algorithms. 

%In the context of mesoscopic morphospace which has a 2-D format, convex hull dimension is, at most, $h_{max}=2$, see \textbf{Fig. S1} for further details. Only the first three cases can occur (since $\max(h)$ equals to the ambient space dimension) although, in general, higher-dimensional morphospace can accommodate the notion of volume.

\subsection{Network Configural breadth - A definition}
Recall that, in the main text, we define the equivalent notion of configural breadth using functional reconfiguration and preconfiguration for a given FN. 
\[
\mathcal{F}_i = f(\mathcal{P}^{FN}_{i}, \mathcal{R}^{FN}_{i})
\]
where $\mathcal{P}^{FN}_{i}$ and $\mathcal{R}^{FN}_{i}$ represent functional preconfiguration and reconfiguration, respectively.
\subsection{Functional Reconfiguration}
In the main text, we address that once the points are well-defined to represent tasks per each functional community, we need now the notion that highlights subject capacity to exploring this cognitive space. We provide a deeper analysis of the drive behind the usage of volume of the convex hull. 
\begin{equation*}
	\mathcal{R}^{FN}_{i} = \mathbf{Vol} (Conv(W^{FN}_i));
\end{equation*}
First of all, since we can only obtain finite number of tasks (hence, points in this space), we see that convex hull notion is logical to represent distinct points (FN tasks) that constitute the Pareto front (hull boundary). To measure the notion of capacity (potential to shift), one needs to measure the notion of spreading given finite number of points in the hull. If we use first order measurements such as distance among two points in the hull, we face the following problems:
\begin{itemize}
	\item inability to capture the reservoir defined by the interior of the convex hull;
	\item assumption of linearity between task points
\end{itemize}
The notion of distance does not cover the space of possibility \cite{avena2015network} parameterized by \textbf{TE} and \textbf{EE}. Hence, second order measurement, i.e. volume (or area in this case), is more appealing. 
\subsection{Functional Preconfiguration}
Analogously, once the points are well-defined in this space, in order to effectively measure the notion of functional preconfiguration, we need to highlight the functional readiness, from a cognition standpoint, to switch between resting configuration to a generic task. Here, we first provide the formula proposed in main-text for functional preconfiguration:
\[
\mathcal{P}^{FN}_{i} = || Rest^{FN}_i - \eta_{W^{FN}_i}||_2
\]
where $Rest^{FN}_i$ and $\eta_{W^{FN}_i}$ represent FN coordinate at rest, and geometric centroid considering all FN tasks. 

Firstly, the geometric centroid of all FN task coordinates might or might not be cognitively possible, i.e. there might not be a connectome that result in FN task centroid being numerically exact. However, that is not the purpose of using this notion. If the goal is to reflect the degree of functionally readiness between resting and task-engagement, the notion of distance, in this case, is meaningful. Here, complexity of trajectory between rest and task-evoked condition is irrelevant to consider.     
\subsection{Subject Sensitivity}
\begin{itemize}
	\item \textbf{Input Matrix}: To quantify subject sensitivity (through configural breadth), for each subject/scan, we obtain one measure. We then concatenate the data into a 100 by 2 matrix and run intra-class correlation (ICC) analysis;
	
	\item \textbf{Null Model Description}: To test subject sensitivity result robustness, for each functional network's preconfiguration or reconfiguration, we keep one column of ICC input intact (Test) and shuffle the second column (Retest) and measure ICC for each permutation. The same procedure is repeated 10,000 times and the 95\%-ile is reported in the main text.
\end{itemize}
\begin{figure*}[h]
	\centering
	{\includegraphics[width=\columnwidth]{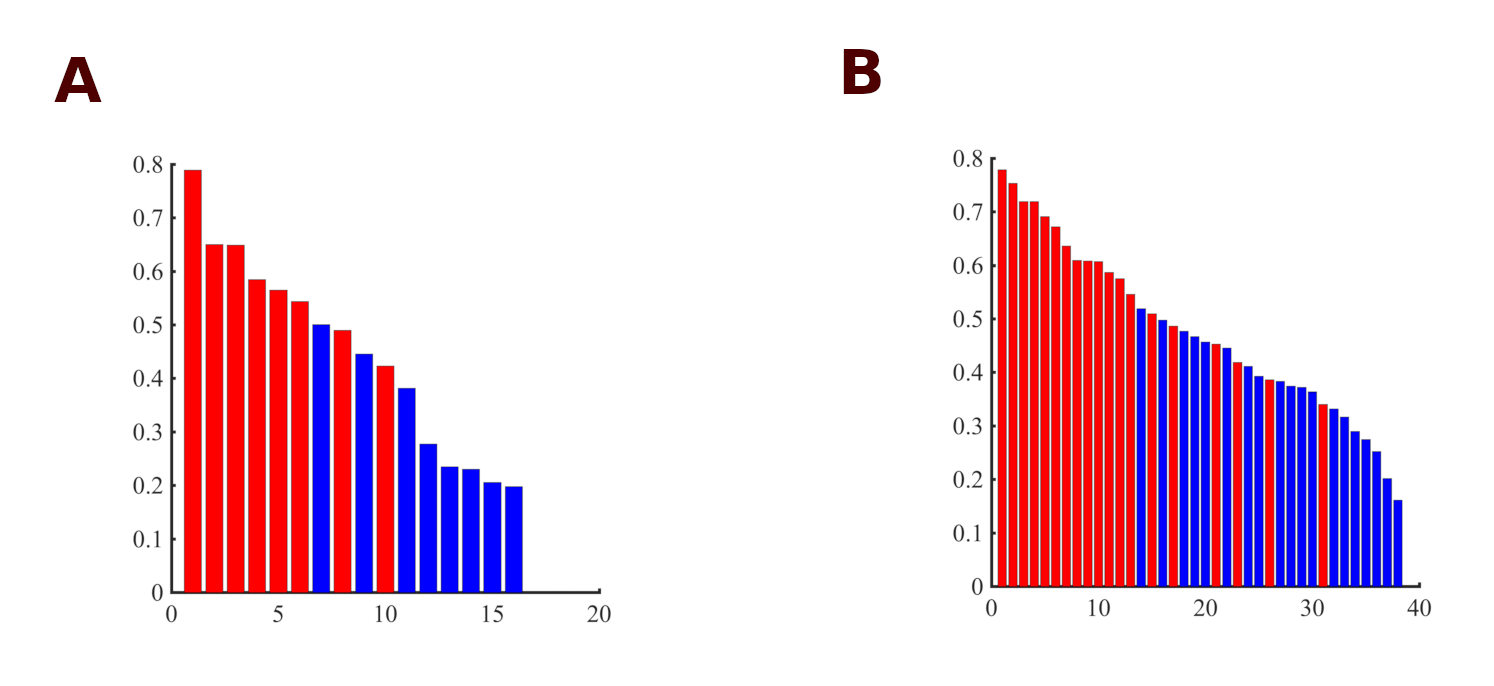} \label{fig:configural breadth_subject_specificity}}
	\caption{\textbf{Network Configural breadth - Subject Specificity Analysis}: FN's pre- (Color: Red) and re- (Color: blue) configuration terms' subject specificity are measured using intra-class correlation. Panel (A) and (B) represent Yeo and colleagues' seven and seventeen FN parcellation \cite{yeo2011organization}, respectively. In both parcellations, FN preconfigurations, overall, tend to have higher subject fingerprints than reconfiguration.}
\end{figure*}
\section{Behavioral Measure Analysis}
\subsection{Iterative Multi-Linear Regression Model (MLM)}
\subsubsection{Model Description}
We apply iteratively multi-linear correlation models (MLM) to correlate $\mathcal{F}_i=f(\mathcal{R}^{FN}_{i}, \mathcal{P}^{FN}_{i})$ with various behavioral measures, $\mathbb{m}_i$. We hypothesize that highly subject sensitive predictor, as described in \textbf{Fig. 7C} (main text) should be prioritized in MLM model. Iteratively, we start by using only 1 predictor ($\mathcal{P}^{FP}$); in every subsequent step, we append one extra predictor to the existing one(s), again, accordingly per panel \textbf{C} of \textbf{Fig. 7} (main text). At the end of iterative process, we consequently obtain 16 MLMs. 
\subsubsection{Optimal MLM - A selection process}
In order to pick the best MLM (and their corresponding number of linear descriptors in the model), we use the model with smallest p-value among all 16 MLMs. 
\subsection{Model Specificity (MS)}
Constructing the MLM to infer the intrinsic relationship is a necessary but not sufficient if the ultimate goal is to discover if there is a truly robust relationship between network configural breadth and behavioral measures. If there exists such robustness, then there has to be a certain degree of specificity in these models such that only significant correlations are observed when linear predictors are correlated with the true behavioral measures. Specifically, network configural breadth, as mathematically formulated using linear descriptors, must show that it is strongly correlated with a designated measures and not anything else, say a randomized vector. 
\subsubsection{Model Description}
We further test the strength of our hypothesis by splitting available data into two subsets: test and validation set. Specifically, we first extract the optimal number of predictors by applying the procedure described in the main article.
We then proceed with the model specificity by creating 2000 simulations; for each simulation - indexed by $j=\cbrac{1,2,3...,J=2000}$ - we first find a randomized order of indices from 1 to 100, denoted as $\vec{d}$, and divide them into five batches (indexed by $i=\cbrac{1,2,...,I=5}$) of 20 subjects. In other words, each batch of 20 randomly picked subjects, indexed by the set $Q_i$, are used to validate the authenticity of the coefficients proposed by utilizing the remaining 80 unpicked subjects. We see that we recover the permutation of the randomized order vector as follows: $\vec{d}=Q = \cup Q_i$. It is important to note that we use this procedure because it minimizes the chance of picking the same (or highly overlapped) batch of 20 subjects.
For each simulation $j$, in each batch $i$, the remaining 80 subjects are then used to acquire multi-linear correlation model's parameters, denoted as $\vec{\beta}\in \mathbb{R}^{[*]}$ where $[*]$ denotes the optimal MLM driven by procedure described above (Notice that we use the same notation in the main text under \textbf{Fig. 8} as well). These corresponding coefficients are then used to predict the remaining 20 unused data points, indexed by $w\in W_i$, denoted as $\hat{y}$. 
\[
\hat{y}_{w} = \vec{\beta}_{0} + \sbrac{\cbrac{\mathcal{P}^{FN}_{w},\mathcal{R}^{FN}_{w}}^{[*]}}\vec{\beta}
\]
where $\cbrac{\mathcal{P}^{FN}_{w},\mathcal{R}^{FN}_{w}}^{[*]} \in \mathbb{R}^{+,[*]}$ is the $[*]$-tupled vector representing functional preconfiguration, reconfiguration, obeying the descending order of concatenated subject sensitivity in \textbf{Fig. 7C} located in the main text. 
Next, for each batch, we compute the correlation between actual values, $y_w$ with predicted ones, $\hat{y}_{w}$ and record the correlating result, denoted as $R_i,\quad\forall i={1,2,...,I=5}$. Consequently, at each simulation, we obtain 5 values of $R_i$ corresponding to 5 batches. Lastly, for each simulation $j$, the mean and standard deviation of 5 validation models $R_i$'s is obtained
\[
R_{j} = \sum_{i=1}^{I} R_{ij} = \langle R_{:,j}\rangle
\quad \& \quad 
\sigma_j = \sqrt{\frac{\sum_{i=1}^{I} (R_{:,j} - R_j)^2}{I}}
\]
Per Central Limit Theorem, the statistic $R_j\mid \forall j=\cbrac{1,2,...,J=2000}$ is normally distributed, i.e. $R_{j}\sim N(\mu_0,\sigma_0)$. This would create an empirically normal distribution $R_{j} \sim N(\mu_0,\sigma_0)$ such that
\[
\mu_0 = \frac{\sum_{j}\sum_{i} R_{ij}}{I\times J}
\quad \& \quad 
\sigma_0 = \sqrt{\frac{\sum_{j=1}^{J}\sigma^{2}_{j}}{J}}
\]
\subsubsection{MS's null model and paired t-test} 
Similarly to the MLMs, we want to test the authenticity of selected models by testing it against artifacts such as random vectors. The same procedure is applied for the random vector to populate the null model's empirically normal distribution (its means is notated as $\mu_1$): 
$R_{j}^{rand} \sim N(\mu_{1},\sigma_{1})$.
Finally, paired t-tests are applied between the two aforementioned distributions, i.e. $R_j$ and $R_j^{rand}$, to test the capacity of configural breadth predictors towards behavioral measures. Interestingly, given the investigated behavioral measures, all null model empirical distributions have very similar first and second moments, independently on behavioral measures.. 

%\newpage
\bibliography{bibliography}
\bibliographystyle{abbrv}
\eject

\end{document}